\documentclass[aps,prb,twocolumn,showpacs,eqsecnum]{revtex4}
\usepackage{bm}

\usepackage{amsmath}
\usepackage{amssymb}
\usepackage{graphicx}
\usepackage{psfrag}

\newcommand{\mm}{\!-\!}
\newcommand{\beq}{\begin{equation}}
\newcommand{\eeq}{\end{equation}}
\newcommand{\beqar}{\begin{eqnarray*}}
\newcommand{\eeqar}{\end{eqnarray*}}

\newcommand{\ua}{\uparrow}
\newcommand{\da}{\downarrow}

\newcommand{\dm}{\diamond}

\newcommand{\dt}{{\text d}}
\newcommand{\etxt}{{\text e}}

\newcommand{\itxt}{{\text i}}

\newcommand{\ttxt}{{\text t}}

\newcommand{\Ktxt}{{\text K}}

\newcommand{\Ttxt}{{\text T}}

\newcommand{\Hh}{\hat{H}}

\newcommand{\Tc}{\mathcal{T}}

\newcommand{\rtarr}{\rightarrow}

\newcommand{\ch}{\hat{c}}

\newcommand{\hh}{\hat{h}}

\newcommand{\Deh}{\hat{\Delta}}

\newcommand{\s}{{\bf s}}
\newcommand{\qb}{{\bf q}}

\newcommand{\Ab}{{\bf A}}

\newcommand{\Bt}{{\tilde{B}}}

\newcommand{\dg}{\dagger}

\newcommand{\lan}{\langle}
\newcommand{\ran}{\rangle}

\newcommand{\al}{\alpha}

\newcommand{\De}{\Delta}
\newcommand{\la}{\lambda}

\newcommand{\sig}{\sigma}

\newcommand{\vphi}{\varphi}

\newcommand{\e}{\epsilon}

\newcommand{\lt}{\left}
\newcommand{\rt}{\right}
\newcommand{\tr}{\text{tr}}

\newcommand{\Ec}{{\cal{E}}}

\begin{document}
\title{Edge excitations of the canted antiferromagnetic phase of the $\nu=0$ quantum Hall state in graphene: a simplified analysis}
\author{Maxim Kharitonov}
\affiliation{
Center for Materials Theory,
Rutgers University, Piscataway, NJ 08854, USA}
\date{\today}
\begin{abstract}
We perform a simplified analysis of the edge excitations of the canted antiferromagnetic (CAF) phase of the $\nu=0$ quantum Hall state
in both monolayer and bilayer graphene.
Namely, we calculate, within the framework of quantum Hall ferromagnetism,
the mean-field quasiparticle spectrum of the CAF phase
neglecting the modification of the order parameter at the edge.
We demonstrate that, at a fixed perpendicular component $B_\perp$ of the magnetic field,
the gap $\De_\text{edge}$ in the edge excitation spectrum
gradually decreases upon increasing  the parallel component $B_\parallel$,
as the CAF phase continuously transforms to the fully spin-polarized ferromagnetic (F) phase.
The edge gap closes completely ($\De_\text{edge}=0$) once the F phase,
characterized by gapless counter-propagating edge excitations, is reached
at some finite $B_\perp$-dependent value $B_\parallel^*$ and remains closed upon further increase of $B_\parallel$.
This results in an gradual insulator-metal transition,
in which the conductance $G \sim (e^2/h) \exp(-\De_\text{edge}/T)$ grows exponentially with $B_\parallel$
in the range $0<B_\parallel<B_\parallel^*$, while in the gapped CAF phase,
and saturates to a metallic value $G \sim e^2/h$
in the F phase at $B_\parallel>B_\parallel^*$.
This unique transport feature  of the CAF phase
provides a way to identify and distinguish it from other competing phases
of the $\nu=0$ quantum Hall state in a tilted-field experiment.

\end{abstract}
\maketitle

\section{Introduction}

Current transport experiments~\cite{Du,Bolotin,Dean,Ghahari,Feldman,Martin,Zhao,Weitz,Mayorov,Freitag,Velasco}
provide compelling evidence for the interaction-induced nature of the ground states
in monolayer (MLG) and bilayer (BLG) graphene.
The most robust of them are observed in the quantum Hall regime
at integer filling factors $\nu$ corresponding to partially filled Landau levels (LLs).
Most commonly, such states belong to the class of
the so-called quantum Hall ferromagnets (QHFMs)~\cite{NM,YDM,AF,Goerbig,JM,Nomura,Barlas,APS,NL3,GorbarBLG,MKMLG,MKBLG,QHFMnote} --
bulk-incompressible states with spontaneously broken symmetry in the valley-spin space.

The key physical challenge related to these states is identifying how exactly the symmetry is broken in a real system.
One of the most intriguing questions concerns the nature
the $\nu=0$ quantum Hall state with half-filled zero-energy LL, realized at the charge neutrality point.
Its characteristic experimental signature in both MLG and BLG is the highly insulating behavior~\cite{Du,Bolotin,Dean,Ghahari,Feldman,Martin,Zhao,Weitz,Mayorov,Freitag,Velasco,YZhang,Checkelsky,LZhang}
of the two-terminal or Hall-bar longitudinal conductance,
a strong indication that both bulk and edge charge excitations of the state are gapped.

\begin{figure*}
\centerline{
\includegraphics[width=.33\textwidth]{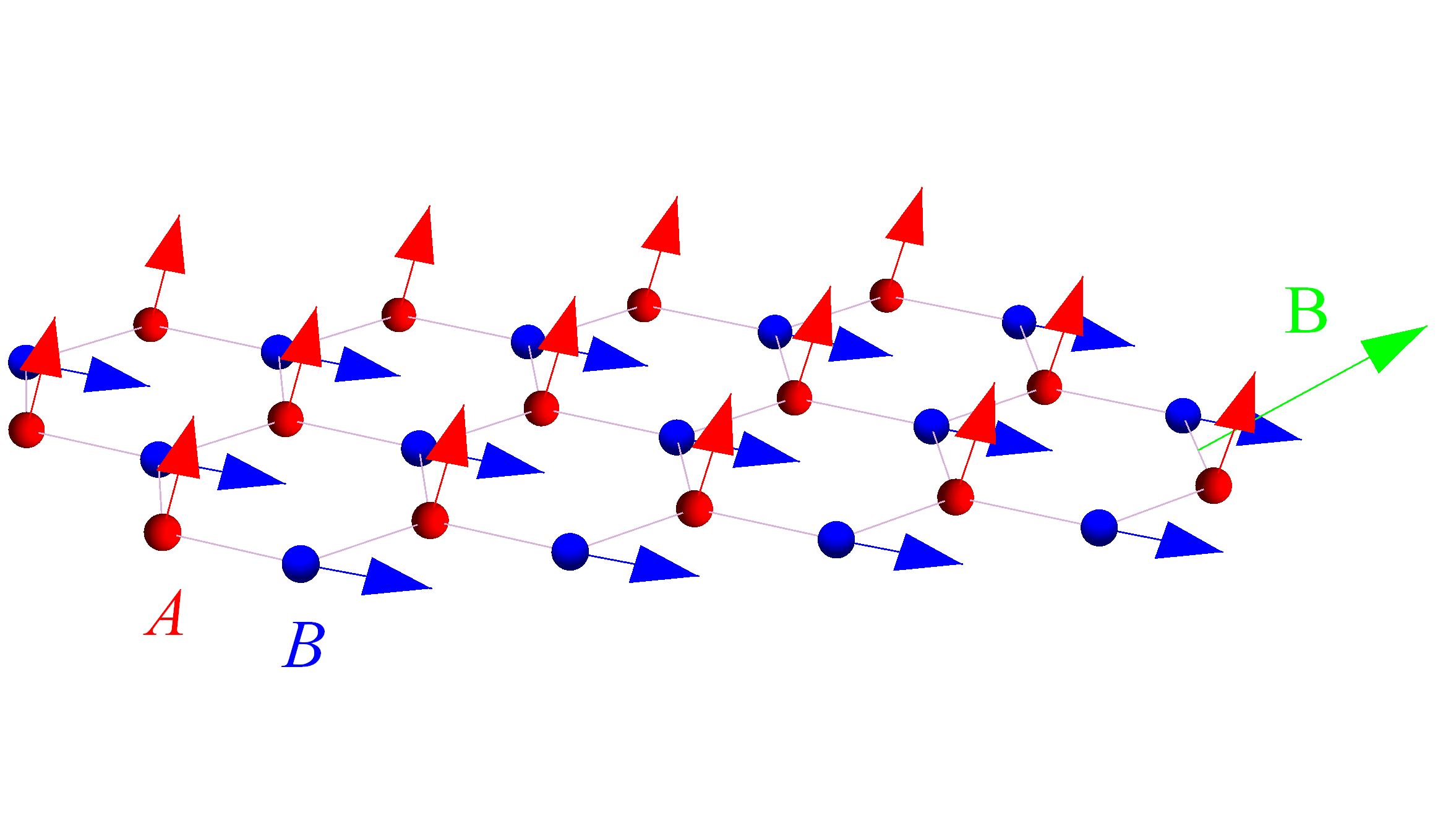}
\includegraphics[width=.33\textwidth]{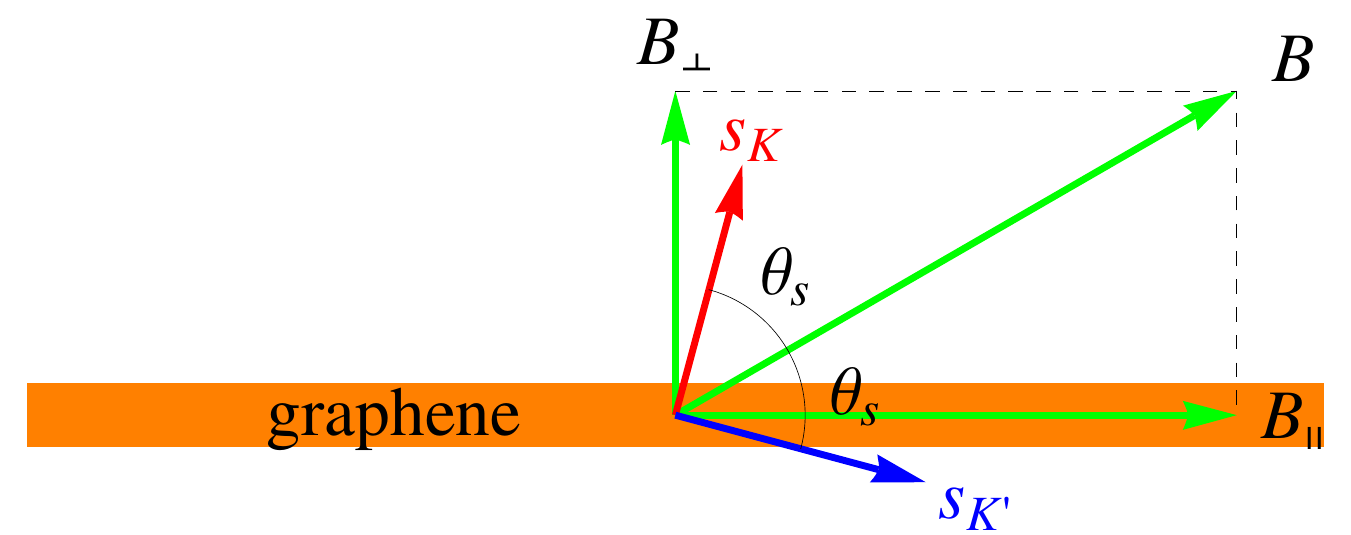}
\includegraphics[width=.33\textwidth]{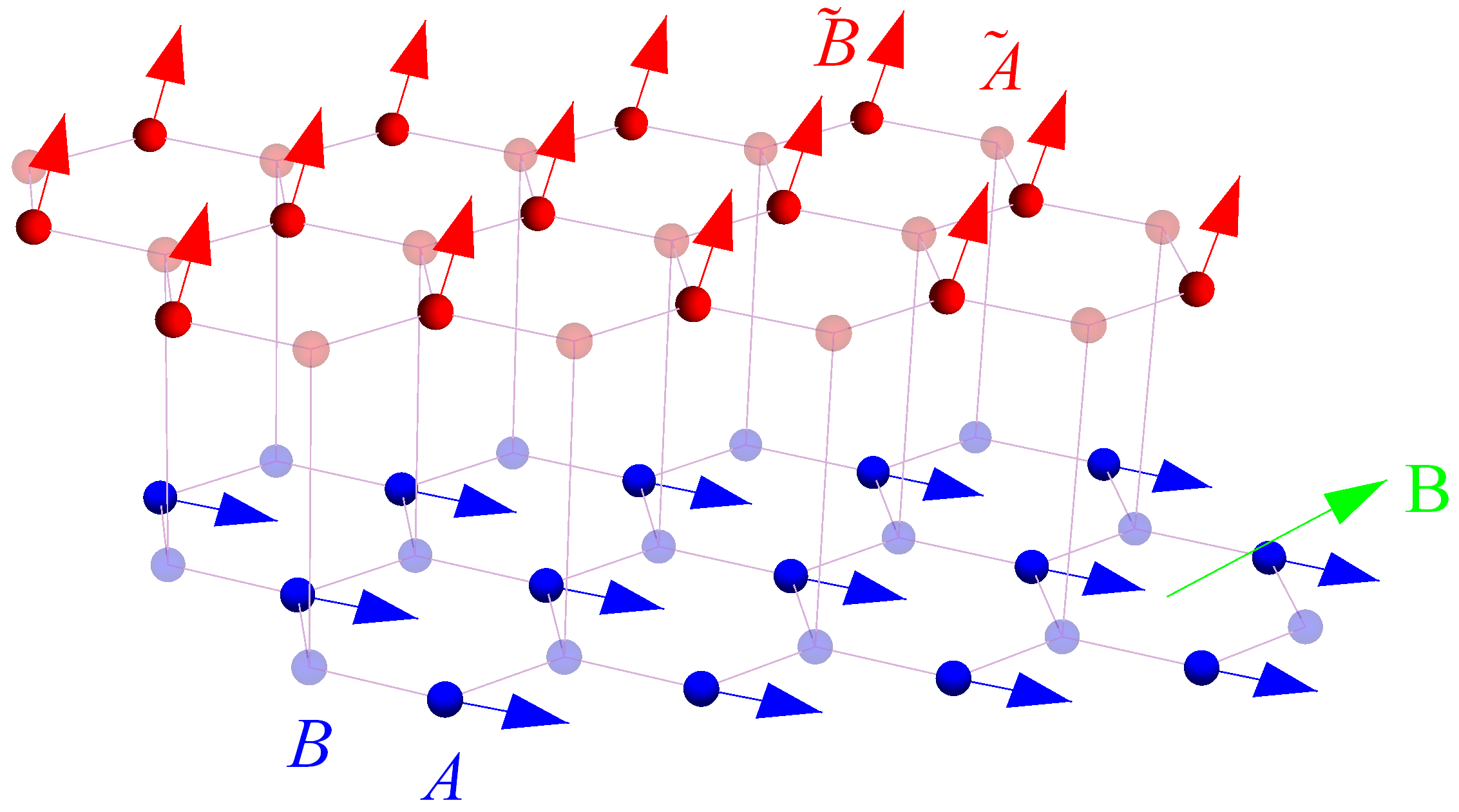}}
\caption{(Color online) 
Canted antiferromagnetic (CAF) phases of the $\nu=0$ quantum Hall state in monolayer (MLG) (left) and bilayer (BLG) (right) graphene.
The spin polarizations $\s_K$ and $\s_{K'}$ of the valleys=sublattices in MLG and valleys=sublattices=layers in BLG
have equal projections on the direction of the total magnetic field (chosen as the $z$ axis in the spin space)
and are antiparallel in the perpendicular plane.
In each valley, $K$ and $K'$,
the wave-functions of the zero-energy Landau level ($\e=0$ LL) reside on either one of the sublattices, $A$ or $B$, in MLG,
and on either one of the sublattice, $\Bt$ or $A$, and therefore in either one of the layers in BLG.
Thus for the $\nu=0$ state,
the valley, sublattice, and layer degrees of freedom are equivalent:
$K\leftrightarrow A$, $K'\leftrightarrow B$ in MLG and
$K\leftrightarrow \Bt \leftrightarrow (\text{top layer})$, $K'\leftrightarrow A \leftrightarrow (\text{bottom layer})$ in BLG.
Throughout the paper, we refer to the spin polarizations $\s_K$ and $\s_{K'}$ by their valley indices.
}
\label{fig:CAF}
\end{figure*}

In Ref.~\onlinecite{MKBLG}, a specific conclusion about the nature
of the experimentally realized insulating $\nu=0$ state in BLG was made,
namely, that it is a canted antiferromagnetic (CAF) phase of the $\nu=0$ QHFM,
in which the spin polarizations $\s_K$ and $\s_{K'}$ of the valleys=sublattices=layers
have equal projections on the direction of the total magnetic field and are antiparallel in
the perpendicular plane, see Fig.~\ref{fig:CAF} and caption to it.
This conclusion was based on the argument
that CAF is the only phase on the generic phase diagram of the $\nu=0$ QHFM, obtained in Refs.~\onlinecite{MKMLG,MKBLG},
consistent with the transport data of Weitz {\em et al.}~\cite{Weitz} and Velasco {\em et al.}~\cite{Velasco}
on dual-gated BLG devices, specifically, with the observation of the insulator-insulator phase transitions
in the perpendicular electric field.

Experimental verification of this conclusion
requires concrete theoretical predictions for measurable quantities,
which would allow one to distinguish the CAF phase from other potential candidates.
In this regard, the CAF phase of the $\nu=0$ state in both MLG and BLG is expected
to exhibit a unique transport property in the tilted magnetic field.
This property concerns the edge charge excitations of the CAF phase and
was  anticipated in Ref.~\onlinecite{MKBLG} based on
the current understanding~\cite{Abanin,FB,Gusynin,JM} of the edge excitations of the antiferromagnetic (AF) and
fully spin-polarized ferromagnetic (F) phases
of the $\nu=0$ state and a general ``by continuity'' argument.

Reiterating this argument,
(i) the CAF phase continuously interpolates between the AF ($\theta_s=\pi/2$) and F ($\theta_s=0$) phases,
as the angle $2\theta_s$ between the spin polarizations $\s_{K,K'}$ of the valleys=sublattices=layers
is varied, $\s_K \s_{K'} = \cos 2 \theta_s$, Fig.~\ref{fig:CAF};
(ii) according to the existing studies~\cite{Abanin,FB,Gusynin,JM},
the AF and F phases %[are expected to][should]
have gapped and gapless %(counterpropagating)
edge charge excitations,
respectively (note that, at the same time, the bulk charge excitations of any phase of a generic QHFM are gapped);
(iii)
therefore, by continuity, upon decreasing $\theta_s$,
as the CAF phase transforms to the F phase,
the gap $\De_\text{edge}(\theta_s)$ of the edge charge excitations
of the CAF phase has to {\em gradually decrease}
and close completely once the F phase is reached, $\De_\text{edge}(\theta_s=0)=0$.

The optimal angle $2 \theta_s$
between the spin polarizations is controlled~\cite{MKMLG,MKBLG}
by the ratio of the Zeeman energy $\e_Z=\mu_B B$, dependent on the total magnetic field $B=\sqrt{B_\perp^2+B_\parallel^2}$,
and the valley ``isospin'' anisotropy energy $u_\perp=u_\perp(B_\perp)$,
dependent on the field component $B_\perp$ perpendicular to the sample:
\[
    \cos \theta_s = \frac{\e_Z}{2 |u_\perp|}.
\]
The anisotropy energy $u_\perp$, defined in Secs.~\ref{sec:bulk} and \ref{sec:BLG},
originates from the Coulomb or electron-phonon interactions at the lattice scale, see Ref.~\onlinecite{MKMLG} for details.
As the key properties,
$u_\perp$ is
(i) linear in $B_\perp$, $u_\perp \sim e^2 a /l_B^2 \sim 1-10 B_\perp [\Ttxt]\Ktxt$, ($a$ is some lattice spatial scale, $l_B =\sqrt{c/(e B_\perp)}$ is
the magnetic length, and we set $\hbar=1$ throughout the paper except for the conductance values $e^2/h$),
if the critical renormalizations~\cite{AKT,MKMLG,Lemonik,Vafek,Lemonik2,Cvetkovic} are weak,
or/and (ii) much greater than the Zeeman energy $\e_{Z\perp} = \mu_B B_\perp $ for perpendicular field orientation,
if renormalizations are substantial.

Therefore, in practice, the ratio $\e_Z/|u_\perp|$ can be efficiently changed and the transition from the CAF to F phase
realized by tilting the magnetic field relative to the sample plane, Fig.~\ref{fig:CAF}.
An unambiguous demonstration of the above predicted behavior
of the edge gap $\De_\text{edge}$ of the CAF phase
requires following the evolution of the system with varying the parallel field component $B_\parallel$ at a fixed $B_\perp$.
This way the $B_\perp$-dependent
correlation energies will not change in the process
and the behavior of the CAF phase will be contrasted to that of
the competing spin-singlet charge-density-wave or Kekul\'{e} phases in MLG and
fully layer-polarized or interlayer-coherent phases in BLG,
whose bulk and edge gaps should not be sensitive (much, if at all) to the Zeeman effect.

Thus, reexpressed in practical terms,
at a fixed $B_\perp$,  the edge gap $\De_\text{edge}$ of the CAF phase will gradually decrease upon increasing $B_\parallel$,
as the CAF phase continuously transforms to the F phase.
The edge gap will close completely
at CAF-F phase transition point (which is of the second order at zero temperature)
at some $B_\perp$-dependent finite value $B_\parallel^*$, determined from the condition
\[
    \e_Z = 2|u_\perp| \mbox{ $\Leftrightarrow$ } \mu_B \sqrt{B_\perp^2+B_\parallel^{*2}} = 2|u_\perp(B_\perp)|.
\]
The edge gap will remain closed, $\De_\text{edge}(B_\parallel\geq B_\parallel^*)=0$,
upon further increase of $B_\parallel>B_\parallel^*$, as the system stays in the F phase.

This behavior will manifest itself in the two-terminal or Hall-bar longitudinal conductance $G$
as a gradual insulator-metal transition upon applying $B_\parallel$.
While in the CAF phase at $0 \leq B_\parallel < B_\parallel^*$ ($\e_Z < 2 |u_\perp|$),
the conductance should follow the Arrhenius activation law $G \propto (e^2/h) \exp(-\De_\text{edge}/T)$
determined by the edge gap $\De_\text{edge}$
(provided the contribution from the bulk with a larger gap $\De_\text{bulk}>\De_\text{edge}$, nearly insensitive to $B_\parallel$, is negligible)
and exhibit exponential sensitivity to $B_\parallel$.
Once the F phase is reached and upon further increase of $B_\parallel$, at $B_\parallel \geq B_\parallel^*$ ($\e_Z \geq 2 |u_\perp|$),
the conductance will saturate to metallic values $G\sim e^2/h$ due to conducting channels provided by the gapless counter-propagating edge excitations~\cite{Abanin,FB,Gusynin,JM} of the F phase.
Ideally, if backscattering of the edge modes is negligible, the conductance in the F phase should be quantized
as $G = 2e^2/h$ in MLG and $G=4e^2/h$ in BLG, according to one and two channels per edge, respectively;
in the two-terminal conductance, an extra factor of 2 arises from two edges, while in the
Hall-bar longitudinal conductance it is due to the mode equilibration in the contacts.
If partial backscattering is present,
$G$ will be lower and conductance fluctuations with varying $B_\parallel$ or other parameters can be expected.

In this paper, we substantiate the above expectations by explicitly calculating the edge charge excitations
of the CAF phase of the $\nu=0$ state in both MLG and BLG within a simplified approach.
The majority~\cite{Abanin,Gusynin,JM,Goerbigedge} of the existing works on the edge excitations of the $\nu=0$ state in MLG
neglect the modification of the bulk order parameter at the edge
and calculate the mean-field quasiparticle spectrum.
It was realized by Fertig and Brey~\cite{FB}, on the other hand,
who studied the edge excitations of the F phase
in MLG, that the bulk order cannot be sustained at the edge due to the
emergence of the finite kinetic energy.
They showed that, in fact, a ``domain wall'' is formed between the bulk and edge orders,
where in the latter electrons fully fill the hole branches of the edge spectrum.
The proper lowest-energy edge charge excitations are then
deformed configurations of the domain-wall texture in the valley-spin space
that carry nonzero topological=electric charge.
These excitations are of the same physical nature as the bulk skyrmions~\cite{AF,YDM,APS},
but have a crucially different energetics.

As the comparison of the findings of Refs.~\onlinecite{Abanin,Gusynin,JM} and Ref.~\onlinecite{FB} shows, however,
both approaches predict gapless edge excitations for the F phase.
This suggests that, even though the former approach is not rigorous,
its results would qualitatively agree with those of the latter for other phases, as well.
For this reason and since the approach of Ref.~\onlinecite{FB} is technically considerably more sophisticated,
in this paper, in order to pinpoint the key physics,
we follow the simplified approach of Refs.~\onlinecite{Abanin,Gusynin,JM,Goerbigedge} to study the edge excitations of the CAF phase of the $\nu=0$ state.
Namely, we neglect the modification of the order parameter at the edge and calculate the mean-field quasiparticle excitations,
and for a specific class of ``armchair-like'' boundaries.
The analysis of the problem based on the generalization of the approach~\cite{FB} of Fertig and Brey
will be presented elsewhere~\cite{MKprep}.

We study the problem within the framework of QHFMism~\cite{NM,YDM,AF,Goerbig,JM,Nomura,Barlas,APS,NL3,GorbarBLG,MKMLG,MKBLG}.
Before we proceed, we mention that
an alternative to QHFMism approach to the $\nu=0$ state in MLG called ``magnetic catalysis''
was developed in Refs.~\onlinecite{MC01,MC02,MC1,Herbut,HerbutCAF,MC2}.
While the approaches are arguably different, certain overlap between the results based on the two can be traced.
In particular, the CAF phase of the $\nu=0$ state in MLG
is predicted to undergo a similar evolution in the tilted magnetic field
within both the magnetic catalysis~\cite{HerbutCAF} and QHFMism~\cite{MKMLG} formalisms.

The rest of the paper is organized as follows.
In Secs.~\ref{sec:model}-\ref{sec:excitations}, the mean-field excitations
of the CAF phase of the $\nu=0$ state are studied for the case of MLG.
In Secs.~\ref{sec:model} and \ref{sec:edge}, the model Hamiltonian is presented.
In Sec.~\ref{sec:bulk}, the mean-field bulk ground state is obtained.
In Sec.~\ref{sec:excitations}, the mean-field excitations of the CAF phase are obtained and
their key properties are studied.
In Sec.~\ref{sec:BLG}, the findings of Secs.~\ref{sec:model}-\ref{sec:excitations}
are generalized to the case of BLG.
Concluding remarks are presented in Sec.~\ref{sec:conclusion}.

\section{Hamiltonian for the zero-energy Landau level in monolayer graphene \label{sec:model}}

Due to the formal equivalence~\cite{MKMLG,MKBLG} of the phase diagrams for the $\nu=0$ QHFM in MLG and BLG,
the results for the edge excitations of the CAF phase, as we show here, turns out physically the same as well.
To keep the analysis clear, in Secs.~\ref{sec:model}-\ref{sec:excitations},
we study the case of MLG in more detail and generalize the obtained results to the case of BLG in Sec.~\ref{sec:BLG}.

We start the analysis by writing down the projected Hamiltonian for the $n=0$ LL in MLG,
valid at energies $\e \ll v/l_B$ ($v$ is the Dirac velocity) much smaller than the LL spacing,
\beq
    \Hh=\Hh_0+\Hh_{\itxt\circ} + \Hh_{\itxt \dm} +\Hh_Z,
\label{eq:H}
\eeq
\beq
        \Hh_0 =
        %\pm \sum_p \e(p) c^\dg_{\la \sig, p} \tau^x_{\la\la'} c_{\la'\sig,p}=
        - \sum_p \e(p) \ch^\dg_p \Tc_x \ch_p,
\label{eq:H0}
\eeq
\beq
    \Hh_{\itxt\circ} = \frac{1}{2} \sum_{p_1+p_1'=p_2+p_2'} V^{p_1 p_2}_{p_1' p_2'}
     :\![\ch_{p_1}^\dg \ch_{p_2} ] [\ch_{p_1'}^\dg \ch_{p_2'}]\!:,
\label{eq:Hi0}
\eeq
\beq
    \Hh_{\itxt\dm} = \frac{1}{2} \sum_{\al=x,y,z} g_\al \sum_{p_1+p_1'=p_2+p_2'} \bar{V}^{p_1 p_2}_{p_1' p_2'}
     :\![\ch_{p_1}^\dg \Tc_\al \ch_{p_2} ] [\ch_{p_1'}^\dg \Tc_\al \ch_{p_2'}]\!:,
\label{eq:Hidm}
\eeq
\beq
    \Hh_Z = - \e_Z\sum_p \ch_p^\dg S_z \ch_p.
    %\Hh_V = - \e_V\sum_k \ch_k^\dg \Tc_z \ch_k,
\label{eq:HZ}
\eeq

We consider a half-infinite sample occupying the $x<0$ half-plane
and work in the basis of the bulk single-particle eigenstates
\beq
    |p \la \sig\ran = |p\la \ran \otimes |\sig\ran, \mbox{ $\la=K,K'$ and $\sig=\ua,\da$, }
\label{eq:bulkstates1}
\eeq
\beq
    |pK\ran =(\psi_p, 0 , 0 , 0 ),\mbox{ } |p K'\ran=(0, 0 , 0 , \psi_p),
\label{eq:bulkstates2}
\eeq
\beq
    \psi_p = \frac{\etxt^{\itxt p y}}{\sqrt{L_y}} \frac{\exp\lt[-\frac{(x-x_p)^2}{2 l_B^2}\rt]}{\sqrt[4]{\pi l_B^2}} ,
    \mbox{ } x_p =p l_B^2,
\label{eq:bulkstates3}
\eeq
of the $n=0$ LL in the Landau gauge $\Ab=(0,B_\perp x, 0)$.
The components of the wave-functions $|p\la\ran$ are ordered as $(\psi_{KA}, \psi_{KB}, \psi_{K'A}, \psi_{K'B})$
in the $KK'\otimes AB$ valley-sublattice space.
The states $|p\la\sig\ran$ are characterized by a conserved one-dimensional momentum $p$
along the edge and definite valley $\la$
and spin $\sig$ quantum numbers.
In each valley, $K$ or $K'$, the wave-functions $|p\la\sig\ran$ reside on either one of the sublattices, $A$ or $B$.

For compactness, in Eqs.~(\ref{eq:H0})-(\ref{eq:HZ}),
we arrange the annihilation operators $\ch_{p\la\sig}$ of electrons in the states $|p\la\sig\ran$ into the spinors
\[
    \ch_p=(\ch_{p K\ua},\ch_{p K\da},\ch_{p K'\ua},\ch_{p K'\da})^\ttxt
\]
in the direct product $KK'\otimes s$ of the valley ($KK'$) and spin ($s$) spaces;
$\Tc_\al = \tau_\al^{KK'} \otimes \hat{1}^s$ and $S_z=\hat{1}^{KK'}\otimes \tau_z^s$ are the valley ``isospin'' and real spin operators, respectively,
and $:\ldots:$ denote normal ordering of operators.

Strictly speaking, the single-particle basis of the bulk states $|p\la\sig\ran$
breaks down at the edge. As we show shortly in Sec.~\ref{sec:edge}, however, an important technical convenience
is that, at energies $\e \ll v/l_B$ of relevance to the low-energy theory of the $\nu=0$ QHFM,
the effect of the edge can be incorporated perturbatively
within the basis of the bulk eigenstates $|p\la\sig\ran$,
and with minimal assumptions about the edge properties.
For a specific class of ``armchair-like'' boundaries we consider in this paper,
this leads to the kinetic energy Hamiltonian $\Hh_0$ of the form (\ref{eq:H0}).

As for the remaining terms in $\Hh$,
$\Hh_{\itxt \circ}$ describes
the valley-symmetric screened Coulomb interactions,
$\Hh_{\itxt \dm}$ describes the valley-asymmetric channels of the interactions,
and $\Hh_Z$ describes the Zeeman effect.
The spin quantization axis $z$ is chosen along the total magnetic field, which
can have arbitrary orientation relative to the sample, see Fig.~\ref{fig:CAF}.

The asymmetric interactions $\Hh_{\itxt \dm}$
arise from the actual Coulomb interactions at the lattice scale or electron-phonon interactions with the optical phonon modes
and may be taken as point in the real space.
They are generically characterized by two signed coupling constants $g_\perp\equiv g_x=g_y$ and $g_z$,
whose bare values scale as $g_\al^{(0)} \sim e^2 a $
and can undergo critical renormalizations~\cite{AKT,MKMLG,Lemonik,Vafek,Lemonik2,Cvetkovic}.
Explicitly breaking the valley symmetry, the asymmetric interactions play
a crucial role~\cite{AF,Goerbig,JM,Nomura,MKMLG,MKBLG,Doretto,Chamon,FuchsLederer} in selecting the favored ground state order of the $\nu=0$ QHFM.

The standard expressions for the interaction matrix elements
$V^{p_1 p_2}_{p_1' p_2'}$ and $\bar{V}^{p_1 p_2}_{p_1' p_2'}$ are provided in the Appendix.

\section{``Armchair-like'' boundary \label{sec:edge}}

\begin{figure}
\centerline{\includegraphics[width=.35\textwidth]{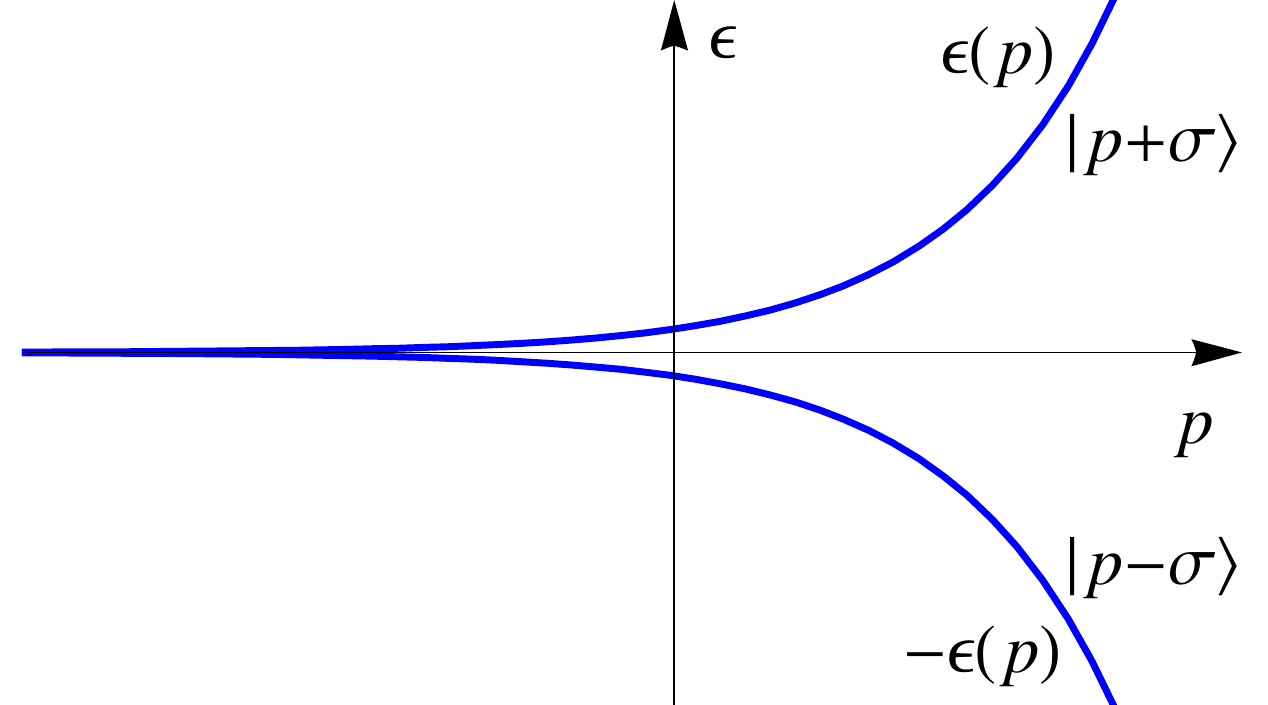}}
\caption{(Color online) The structure of the single-particle spectrum pertaining to the $n=0$ LL
in a MLG sample with ``armchair-like'' boundary, neglecting the Zeeman effect.
}
\label{fig:ep}
\end{figure}

In the effective low-energy Dirac theory for the electron motion in MLG,
valid at energies much smaller than the nearest-neighbor hopping amplitude $t\sim 3\text{eV}$ between the carbon atoms,
an edge of the sample is described by a boundary
condition for the Dirac spinor $\psi=(\psi_{KA}, \psi_{KB}, \psi_{K'A}, \psi_{K'B})^\ttxt$.
As demonstrated in Ref.~\onlinecite{Akhmerov},
under general assumptions of preserved time-reversal and particle-hole symmetries,
a {\em generic edge} of MLG is described by either an armchair or zigzag {\em boundary conditions}.

In this paper, we restrict ourselves to the class of ``armchair-like'' boundaries,
which, by definition, are described by an armchair boundary condition.
Besides the actual armchair edge, more edge structures can belong to this class.
A property of an armchair-like boundary, key to our consideration,
is that it does not contain dispersionless edge states in the absence of the orbital magnetic field.
In the presence of the latter, this translates to the fact that
the number of branches in the single-particle edge spectrum is equal
to the discrete degeneracy of the bulk states $|p\la\sig\ran$, i.e., four.

Let
\[
   |p- \ran = (\psi_{KA},\psi_{KB},\psi_{K'A},\psi_{K'B})
\]
be an exact solution of the boundary problem in a magnetic field~\cite{Abanin,BreyFertig,Gusynin,Mazo} 
for the Dirac equation per given spin projection $\sig$,
characterized by a negative energy $-\e(p)<0$ (hole branch) and pertaining to the $n=0$ LL.
The kinetic energy $\e(p) \rtarr 0$ is flat in the bulk ($p \lesssim 0$) and grows at the edge ($p \gtrsim 0$);
its exact functional, although obtained~\cite{Abanin,BreyFertig,Gusynin,Mazo}, is not essential right now.
The eigenstate
\beq
    |p+ \ran = (\psi_{KA}, -\psi_{KB}, \psi_{K'A}, -\psi_{K'B}).
\label{eq:holestate}
\eeq
with the positive energy $\e(p)>0$ (particle branch)
is obtained from $|p-\ran$ by the particle-hole transformation,
\beq
    \e\rtarr -\e: %\mbox{ } \psi_{\la B} \rtarr -\psi_{\la B} [\mbox{ or }
    \psi_{\la B} \rtarr -\psi_{\la B}, \mbox{ }\la=K,K'
\label{eq:ph}
\eeq

By the assumption of an armchair-like boundary,
there are no other single-particle states besides $|p\pm\ran$ at energies $|\e|\ll v/l_B$.
The second-quantized Hamiltonian for the kinetic energy reads
\beq
        \Hh_0 = \sum_p \e(p) (c^\dg_{p+\sig} c_{p+\sig} - c^\dg_{p-\sig} c_{p-\sig}),
\label{eq:H0pm}
\eeq
where $c_{p\pm\sig}$ are the electron annihilation operators for the states $|p \!\pm\! \sig\ran=|p\pm \ran \otimes |\sig\ran$, with spin included.
The spectrum is shown in Fig.~\ref{fig:ep}.

Deep in the bulk, where the kinetic energy $\e(p)\rtarr 0$ vanishes,
the exact eigenstates $|p\pm\ran$
{\em must} evolve into the linear combinations of the bulk states $|pK\ran$ and $|pK'\ran$ [Eqs.~(\ref{eq:bulkstates2}) and (\ref{eq:bulkstates3})]
that are related by the particle-hole symmetry,
\beq
    |p\pm\ran  = \frac{1}{\sqrt{2}} (|p K\ran \mp |p K'\ran) \mbox{ at } \e(p)\rtarr 0,
\label{eq:pm}
\eeq
The electron operators are related accordingly,
\beq
    c_{p\pm\sig} = \frac{1}{\sqrt{2}}(c_{pK\sig} \mp c_{pK'\sig}) \mbox{ at } \e(p)\rtarr 0.
\label{eq:cpm}
\eeq

Immediately at the edge, where $\e(p) \sim v/l_B$,
the bulk states $|p\la\sig\ran$ [Eqs.~(\ref{eq:bulkstates1})-(\ref{eq:bulkstates3})] are not well-defined.
However, not too close to the edge, where $\e(p) \ll v/l_B$,
one may treat the kinetic energy as a perturbation,
neglecting the modification of the wave-functions but taking into account the energy splitting
of the states $|p\la\sig\ran$.
In this case, one may still use the relation (\ref{eq:cpm}).
Substituting Eq.~(\ref{eq:cpm}) into Eq.~(\ref{eq:H0pm}), we obtain an approximate expression (\ref{eq:H0})
for the kinetic energy in the basis of the bulk states $|p\la\sig\ran$, valid not too close to the edge, so that $\e(p) \ll v/l_B$.

Thus, at energies below the LL spacing $v/l_B$,
the effect of an  armchair-like edge on the $n=0$ LL states
can be taken into account perturbatively
while remaining within the basis of the bulk eigenstates $|p\la\sig\ran$.
The effect amounts to an effective ``Zeeman'' field $-\e(p) (1,0,0)$
along the $x$ direction in the $KK'$-isospin space
that hybridizes the $|pK\sig\ran $ and $|pK'\sig\ran$ states,
favoring the occupation of the states $|p\mm\sig\ran$ of the hole branch, Eqs.~(\ref{eq:holestate}), (\ref{eq:pm}), and (\ref{eq:cpm}).
Note that the direction of this ``Zeeman'' field in the $xy$ isospin plane could be chosen arbitrary
due to the freedom of choice of the phase factor in the particle-hole transformation (\ref{eq:ph}).

\section{Bulk ground state \label{sec:bulk}}

The strongly interacting $\nu=0$ state is described by the theory of QHFMism~\cite{NM,YDM,AF,Goerbig,JM,Nomura,Barlas,APS,NL3,GorbarBLG,MKMLG,MKBLG}.
In this Section, we briefly recover the results of Refs.~\onlinecite{MKMLG,MKBLG} for the bulk ground state,
pertaining to the CAF phase.

One constructs a Slater-determinant state
\beq
    \Psi=  \prod_p
    \lt(\sum_{\la\sig} \lan \la \sig|\chi_a \ran c^\dg_{p\la\sig} \rt)
    \lt(\sum_{\la'\sig'} \lan \la'\sig'|\chi_b\ran  c^\dg_{p\la'\sig'} \rt)|0\ran,
\label{eq:Psi}
\eeq
in which two electrons per each orbital $p$ of the $n=0$ LL occupy
{\em arbitrary} mutually orthogonal states $\chi_{a,b}$ in the $KK' \otimes s $ space.
It is straightforward to show that $\Psi$ is an exact eigenstate of the Hamiltonian (\ref{eq:Hi0})
of the SU(4)-symmetric interactions for any choice of the spinors $\chi_{a,b}$,
\[
    \Hh_{\text{i}\circ} \Psi =E_0 \Psi.
\]

For a wide class of repulsive interactions, one can expect the eigenstates $\Psi$ to be exact ground states
by the Hund's rule argument. This is the main assumption of the QHFMism theory, also employed in this paper.

Thus at the level of symmetric interactions,
the ground state is known exactly, but it is highly degenerate.
This degeneracy is uniquely parameterized by the order parameter matrix
\beq
    P=\chi_a \chi_a^\dg + \chi_b \chi_b^\dg,
\label{eq:P}
\eeq
which satisfies the properties of a projection operator,
\beq
    P^\dg=P, \mbox{ } P^2=P, \mbox{ } \tr P =2.
\eeq

The favored order $P$ is determined by the effects that explicitly breaks the SU(4) symmetry
in the $KK' \otimes s$ space: valley-asymmetric interactions (\ref{eq:Hidm}) and the Zeeman effect (\ref{eq:HZ}).
These effects are taken into account perturbatively by calculating their energy expectation values,
\begin{eqnarray}
    \Ec(P) &=& \Ec_\dm(P) + \Ec_Z(P),
\label{eq:Ec}\\
    \Ec_\dm(P) &=& \lan \Psi| \Hh_{\itxt \dm} |\Psi \ran/N = \nonumber \\
        & = &
        \frac{1}{2}\sum\nolimits_{\al} u_\al \{ \tr^2 [\Tc_\al P ] - \tr [\Tc_\al P \Tc_\al P ] \},
\label{eq:Edm} \\
%    \Ec_V(P) &=& \lan \Psi| \Hh_V |\Psi \ran/N = -  \e_V \,\tr [ \Tc_z P ],\\
    \Ec_Z(P) &=& \lan \Psi| \Hh_Z |\Psi \ran/N = - \e_Z \,\tr [ S_z P ].
\label{eq:EZ}
\end{eqnarray}
Here $N=\sum_p 1$ is the number of orbital states, equal to the number of flux quanta threading the sample.
The asymmetric interactions result in the isospin anisotropy energy $\Ec_\dm(P)$, characterized by two signed energies
\beq
    u_\perp \equiv u_x=u_y = \frac{g_\perp}{2\pi l_B^2}, \mbox{ }
    u_z = \frac{g_z}{2\pi l_B^2}.
\label{eq:u}
\eeq

Minimization of the energy $\Ec(P)$ of the SU(4)-symmetry-breaking effects for arbitrary values of $u_{\perp,z}$ and $\e_Z$,
resulting in the generic phase diagram for the $\nu=0$ QHFM in MLG and BLG, was carried out in Refs.~\onlinecite{MKMLG,MKBLG}.
In this paper, we will be interested in the canted antiferromagnetic (CAF) phase, argued in Ref.~\onlinecite{MKBLG}
to be realized in the insulating $\nu=0$ state of the real BLG.

The CAF phase is realized when the isospin anisotropy $\Ec_\dm(P)$ alone
favors the AF phase. This occurs for $u_z>-u_\perp>0$,
the condition assumed to be satisfied in the rest of paper.
In this case, in the presence of the Zeeman effect,
the energy $\Ec(P)$ is minimized by either the CAF or F phases~\cite{MKMLG,MKBLG},
with $\chi_a=|K\ran \otimes |\s_K\ran$ $\chi_b=|K'\ran \otimes |\s_{K'}\ran$,
in which the spin polarizations
\beq
    \s_{K,K'} = (\pm \sin \theta_s \cos \vphi_s, \pm \sin \theta_s \sin \vphi_s, \cos \theta_s)
\label{eq:s}
\eeq
of the valleys=sublattices in MLG and valleys=sublattices=layers in BLG
have equals projections on the direction of the total magnetic field
and are antiparallel in the perpendicular plane, Fig.~\ref{fig:CAF}.

The energy
\[
    \Ec(P) = -u_z-u_\perp-u_\perp \cos 2 \theta_s - 2 \e_Z \cos \theta_s
\]
of this family of states is minimized at
\beq
    \cos \theta_s = \lt\{ \begin{array}{ll}
        \frac{\e_Z}{2|u_\perp|}, & \e_Z <2 |u_\perp|  \text{ (CAF)},\\
        1, & \e_Z>2 |u_\perp| \text{ (F)}.
        \end{array}
        \rt.
\label{eq:thetas}
\eeq
I.e., for $\e_Z<2|u_\perp|$ the ground state is a CAF phase with the optimal angle $2\theta_s$
between the spins and energy
\[
    \Ec^\text{CAF}= -u_z-\frac{\e_Z^2}{2|u_\perp|},
\]
while for $\e_Z>2|u_\perp|$
it is a fully spin-polarized F phase with the energy
\[
    \Ec^\text{F} = 2 |u_\perp| -u_z.
\]

The CAF phase has a U(1)-degeneracy according to the choice of the spin orientation $\vphi_s$
in the plane perpendicular to the total magnetic field.
For the calculations below, we will assume a specific orientation $\vphi_s=0$,
in which case the order parameter (\ref{eq:P}) of the CAF/F phases takes the form
\beq
    P = \frac{1}{2} \hat{1} \otimes (\hat{1}+\cos \theta_s \tau_z) + \frac{1}{2} \tau_z \otimes \sin \theta_s \tau_x
\label{eq:PCAF}
\eeq

As discussed in the Introduction,
the obtained bulk order parameter (\ref{eq:PCAF})
cannot be sustained at the edge due to the emergence of the finite kinetic energy
and will be necessarily modified at $p$, such that $\e(p) \gtrsim \Ec(P)$.
For the reason mentioned there, however,
in this paper, we will neglect this fact
and use the bulk order parameter $P$ at all momenta $p$, even where the kinetic energy is not
negligible anymore.

\section{Mean-field excitations of the canted antiferromagnetic phase in monolayer graphene\label{sec:excitations}}

\begin{figure*}
\centerline{\includegraphics[width=.33\textwidth]{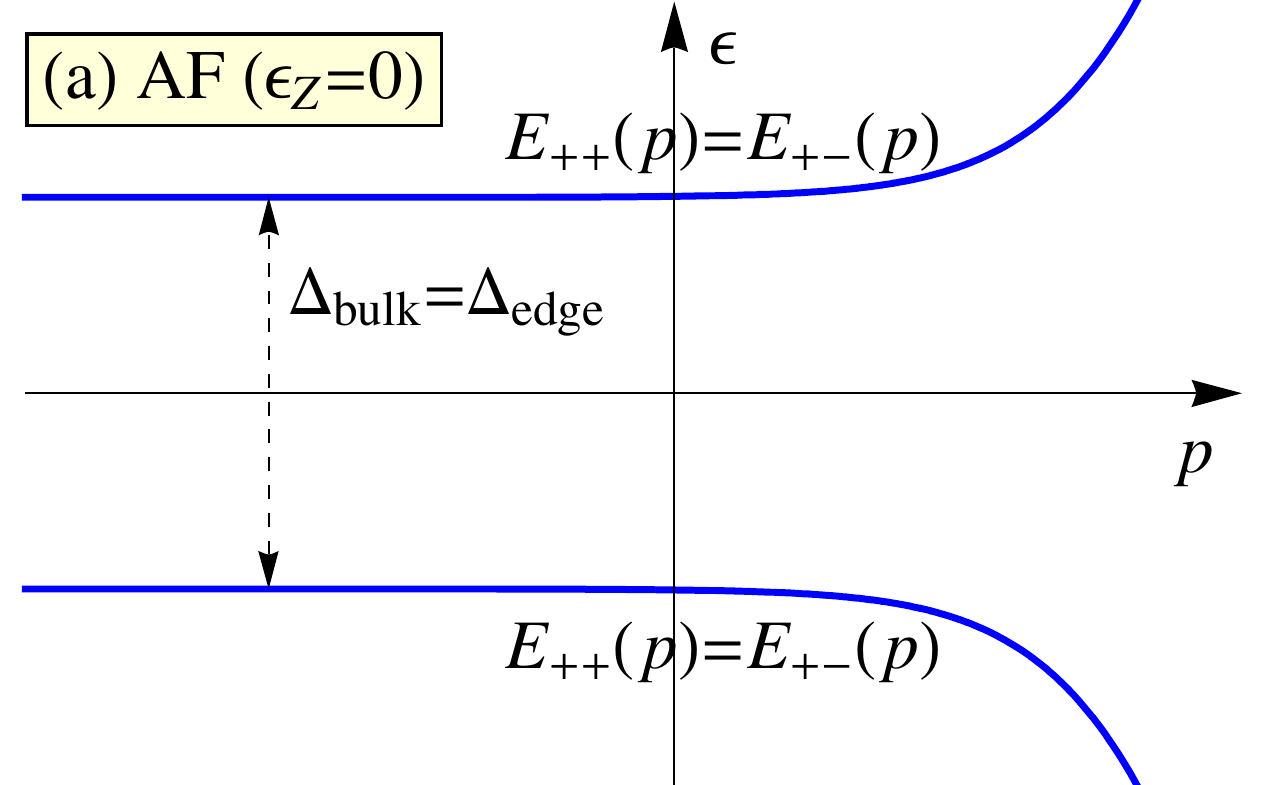}\includegraphics[width=.33\textwidth]{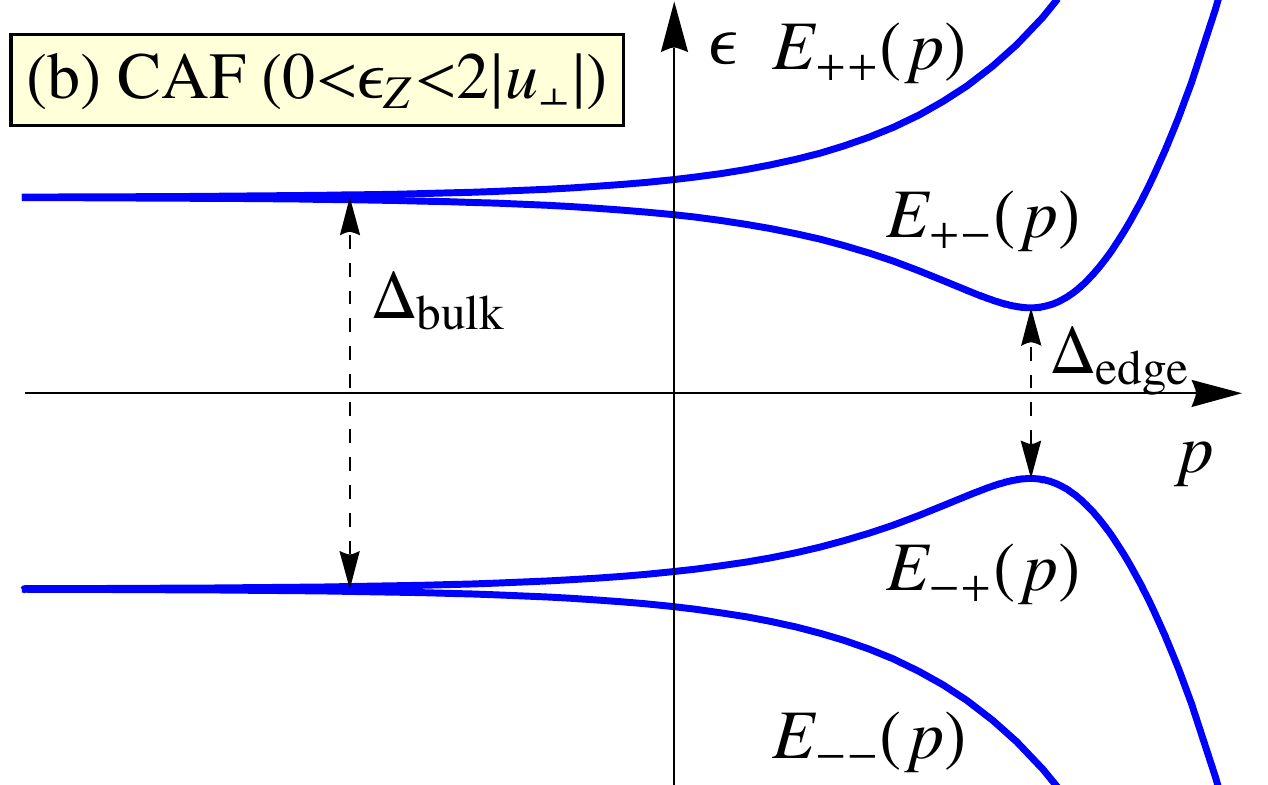}\includegraphics[width=.33\textwidth]{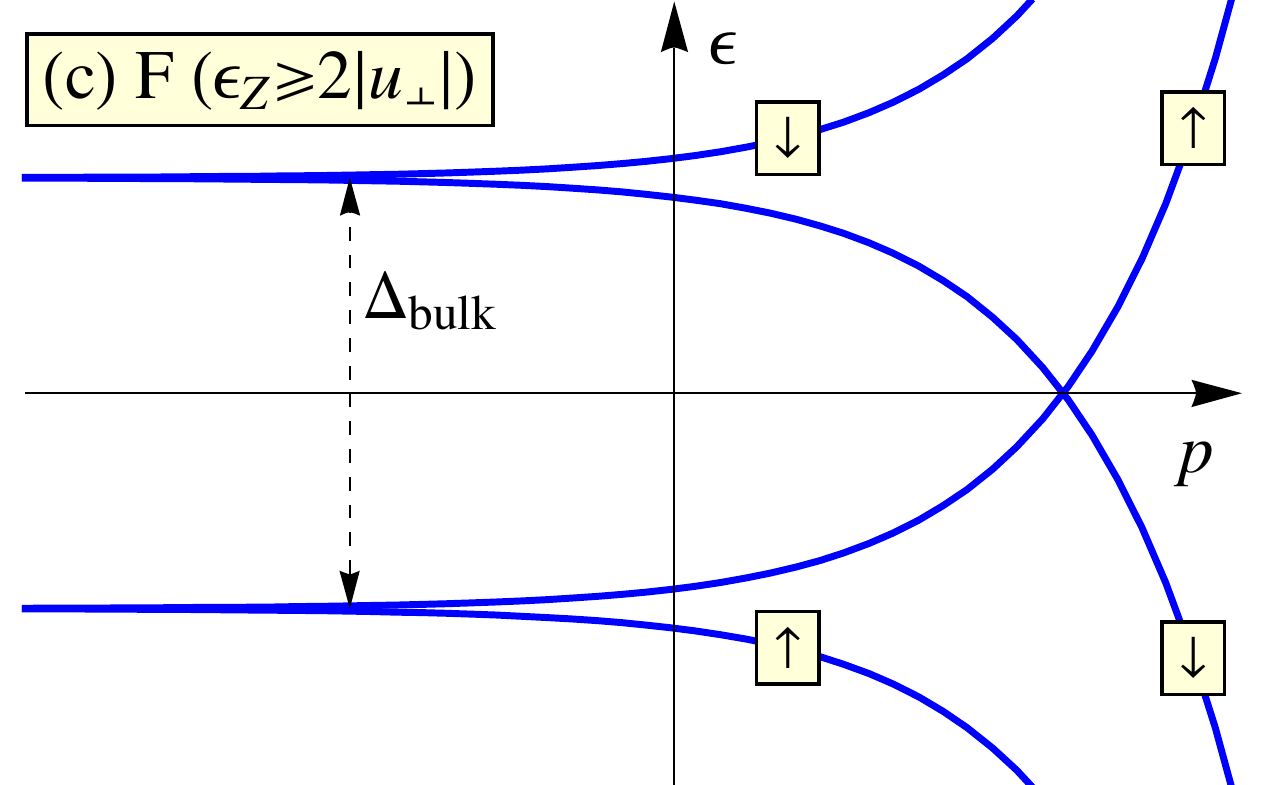}}
\caption{(Color online)
The mean-field spectrum of the (a) antiferromagnetic (AF), (b) canted antiferromagnetic (CAF), and (c) ferromagnetic (F) phases of the $\nu=0$ state
of a monolayer graphene (MLG) sample with armchair-like boundary (see Sec.~\ref{sec:edge} for a definition).
The spectrum consists of four branches $E_{\pm\pm}(p)$ [Eq.~(\ref{eq:Ep})].
The edge gap $\De_\text{edge}$ is determined by the shortest distance between the $E_{+-}(p)$ and $E_{-+}(p) = -E_{+-}(p)$ branches.
The edge gap $\De_\text{edge}$ is maximal in the AF phase (a),
has a smaller value in the CAF phase (b),
and vanishes in the F phase (c),
where gapless counter-propagating edge excitations with opposite spin projections emerge.
The evolution of the edge $\De_\text{edge}$ [Eq.~(\ref{eq:Deedge})] and bulk $\De_\text{bulk}$ [Eq.~(\ref{eq:Debulk})]
gaps with the Zeeman energy $\e_Z$ at fixed interaction energies $u_{0,\perp,z}$
is presented in Fig.~\ref{fig:De}.
}
\label{fig:Ep}
\end{figure*}

\begin{figure}
\hspace{.03\textwidth}\includegraphics[width=.33\textwidth]{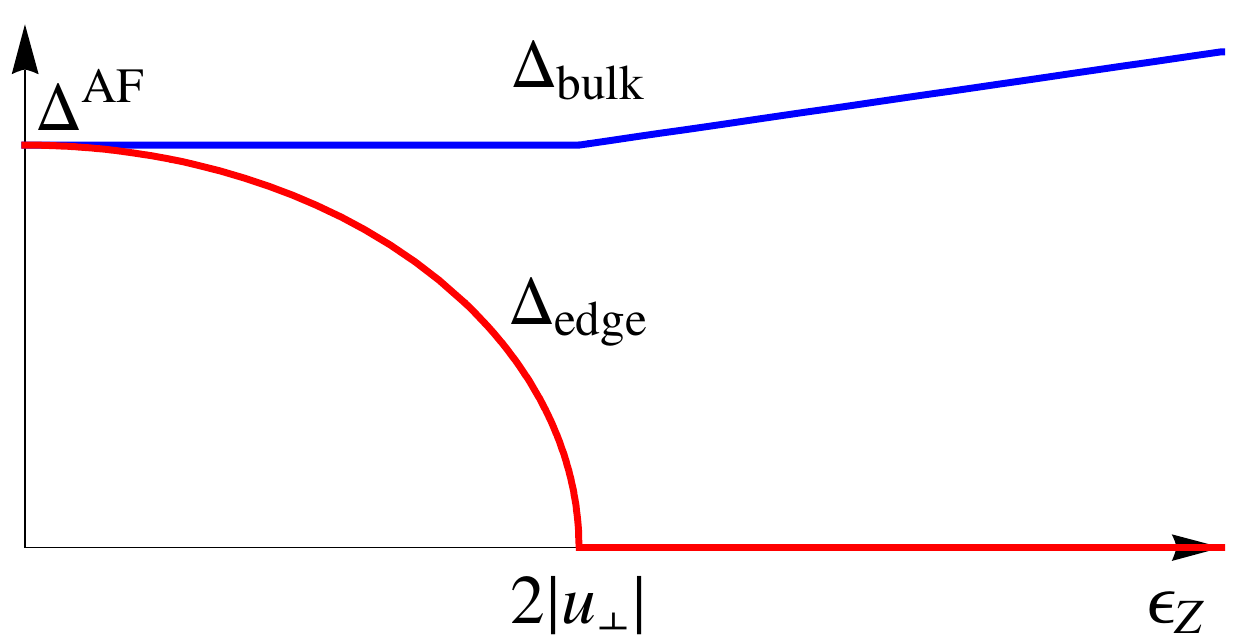}
\vspace{1mm}
\includegraphics[width=.35\textwidth]{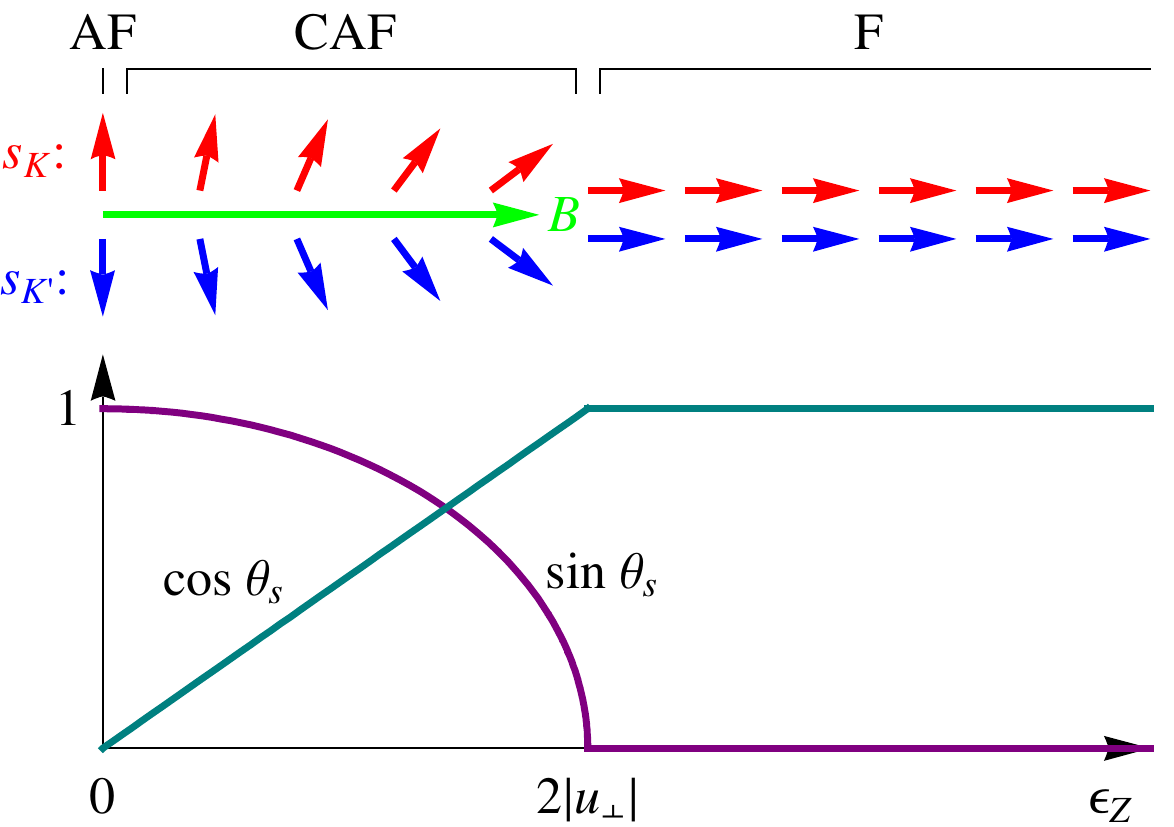}
\caption{(Color online)
(Top) The evolution of the edge $\De_\text{edge}$ [Eq.~(\ref{eq:Deedge})] and bulk $\De_\text{bulk}$ [Eq.~(\ref{eq:Debulk})]
gaps in the mean-field excitation spectrum [Eqs.~(\ref{eq:Ep}) and (\ref{eq:EpBLG}) and Figs.~\ref{fig:Ep} and \ref{fig:EpBLG}]
of the AF/CAF/F phases of the $\nu=0$ state in both MLG and BLG upon changing the Zeeman energy $\e_Z$
at fixed values of the interaction energies $u_{0,\perp,z}$.
In practice, such evolution is realized
by varying a parallel field component $B_\parallel$ of the magnetic field at a fixed perpendicular component $B_\perp$, Fig.~\ref{fig:CAF}.
The edge gap $\De_\text{edge}$, determined by the AF component $\De_{zx}$ of the mean-field potential (\ref{eq:De}),
is maximal in the AF phase at $\e_Z=0$ [Figs.~\ref{fig:Ep}(a) and \ref{fig:EpBLG}(a)],
gradually decreases upon increasing $\e_Z$
in the range $0<\e_Z<2|u_\perp|$, while in the CAF phase [Figs.~\ref{fig:Ep}(b) and \ref{fig:EpBLG}(b)],
and vanishes at the CAF-F phase transition at $\e_Z=2|u_\perp|$ and upon further increase of $\e_Z\geq 2|u_\perp|$,
as the system stays in the F phase [Fig.~\ref{fig:Ep}(c) and Fig.~\ref{fig:EpBLG}(c)].
The bulk gap $\De_\text{bulk}$ is constant
in the AF/CAF phases at $\e_Z<2|u_\perp|$ and grow in the F phase due to the Zeeman effect.
(Bottom) The corresponding evolution of the spin polarizations $\s_{K,K'}$ [Eqs.~(\ref{eq:s}) and (\ref{eq:thetas})] of the valleys=sublattices=layers in the AF/CAF/F phases.
}
\label{fig:De}
\end{figure}

Mean-field excitations are obtained by performing
decoupling of the interactions in Eqs.~(\ref{eq:Hi0}) and (\ref{eq:Hidm}),
\beq
    \Hh_{\text{i}\circ}+\Hh_{\text{i}\dm} \rtarr \Hh_\text{i,mf} =- \sum_p \ch^\dg_p \Deh \ch_p,
\eeq
\beq
    \Deh=u_0 P -\sum_{\al=x,y,z} u_\al(\Tc_\al \tr[P\Tc_\al] -\Tc_\al P\Tc_\al),
\label{eq:Dedef}
\eeq
where $u_0 = \sum_{p'} V^{p'p}_{p p'}$ is the exchange energy
of the symmetric interactions (\ref{eq:Hi0}) (we discard the $P$-independent Hartree energy of
the symmetric interactions)
and the isospin anisotropy energies $u_{\perp,z}$ of the asymmetric interactions (\ref{eq:Hidm})
were defined in Eq.~(\ref{eq:u}).

Inserting Eq.~(\ref{eq:PCAF}) into Eq.~(\ref{eq:Dedef}) and discarding the trivial terms $\propto  \hat{1}\otimes \hat{1}$,
we obtain for the mean-field potential of the AF/CAF/F phases 
\begin{eqnarray}
    \Deh&=& \De_{0z} \hat{1}\otimes\tau_z+\De_{zx} \tau_z \otimes \tau_x,
\label{eq:De}
\\
    \De_{0z}&=&\frac{1}{2}(u_0+u_z-2 |u_\perp|) \cos\theta_s,
\label{eq:De0z}
\\
    \De_{zx}&=&\frac{1}{2}(u_0+u_z+2|u_\perp|) \sin\theta_s.
\label{eq:Dezx}
\end{eqnarray}

The full mean-field Hamiltonian takes the form
\[
    \Hh_\text{mf} \equiv \Hh_0+\Hh_\text{i,mf}+\Hh_Z = \sum_p \ch^\dg_p \hh \ch_p,
\]
\beq
    \hh = -\e(p) \tau_x \otimes \hat{1} - (\e_Z+\De_{0z}) \hat{1}\otimes \tau_z - \De_{zx} \tau_z \otimes \tau_x.
\eeq

The single-particle Hamiltonian $\hh$ is straightforwardly diagonalized. This yields the four branches
\beq
    E_{\pm\pm}(p) = \pm \sqrt{[\e(p) \pm (\e_Z+\De_{0z})]^2  + \De_{zx}^2}
\label{eq:Ep}
\eeq
of the mean-field spectrum of the AF/CAF/F phases of the $\nu=0$ state in MLG,
plotted in Fig.~\ref{fig:Ep}.

At the edge, the branches $E_{+-}(p)$ and $E_{-+}(p)=-E_{+-}(p)$ come closest to each other and the
gap $\De_\text{edge}$ in the edge excitation spectrum is determined by the branch minimum,
$\De_\text{edge} =2 E_{+-}(p_0)$, where $\dt E_{+-}(p)/ \dt p|_{p_0}=0$.
We obtain
\[
    \De_\text{edge} = 2\De_{zx} =
\]
\beq
    = \lt\{ \begin{array}{ll}
        \De^\text{AF}\sqrt{1-(\frac{\e_Z}{2 |u_\perp|})^2}, & \e_Z <2 |u_\perp|  \text{ (CAF)},\\
        0, & \e_Z\geq 2 |u_\perp| \text{ (F)},
        \end{array}
        \rt.
\label{eq:Deedge}
\eeq
where
\beq
    \De^\text{AF}= u_0+u_z+2|u_\perp|
\label{eq:DeAF}
\eeq
is the edge and bulk gap of the AF phase at $\e_Z=0$.

The bulk gap  $\De_\text{bulk} = 2 E_{+\pm}(p \rtarr -\infty)$ is obtained by setting $\e(p)=0$ in Eq.~(\ref{eq:Ep}) and
equals
\[
    \De_\text{bulk} = 2\sqrt{(\De_{0z}+\e_Z)^2  + \De_{zx}^2} =
\]
\beq
    = \lt\{ \begin{array}{ll}
        \De^\text{AF}, & \e_Z < 2 |u_\perp|  \text{ (CAF)},\\
        \De^\text{AF}+2(\e_Z-2|u_\perp|), & \e_Z\geq 2|u_\perp| \text{(F)}.
        \end{array}
        \rt.
\label{eq:Debulk}
\eeq

Equations (\ref{eq:Ep}), (\ref{eq:Deedge}), and (\ref{eq:Debulk}) for the spectrum of the mean-field excitations of the AF/CAF/F phases in MLG
constitute the main result of the work. Below we discuss their key properties
and demonstrate that they do lead to behavior anticipated in the Introduction.

The mean-field potential (\ref{eq:De}) of the CAF phase is a mixture of the ferromagnetic ($\De_{0z}$)
and antiferromagnetic ($\De_{zx}$) components;
their relative value
is fully controlled by the ratio $\e_Z/|u_\perp|$, which determines the angle $2\theta_s$
between the spins polarizations $\s_{K,K'}$ [Eq.~(\ref{eq:thetas})].

As discussed in the Introduction, efficiently changing the ratio $\e_Z/|u_\perp|$ in the experiment requires tilting the magnetic field,
Fig.~\ref{fig:CAF}.
Let us discuss the evolution, plotted in Figs.~\ref{fig:Ep} and \ref{fig:De},
of the edge $\De_\text{edge}$ [Eq.~(\ref{eq:Deedge})] and  bulk $\De_\text{bulk}$ [Eq.~(\ref{eq:Debulk})]
gaps upon changing the Zeeman energy $\e_Z$
at fixed values of the interaction energies $u_{0,\perp,z}$.
Practically, this corresponds to fixing $B_\perp$ and applying $B_\parallel$.

According to Eq.~(\ref{eq:Deedge}), the edge gap $\De_\text{edge}$ is equal to twice the AF component $\De_{zx}$ of the mean-field potential.
In the theoretical limit of vanishing Zeeman energy $\e_Z=0$, the phase is purely AF;
the F component $\De_{0z}=0$ is absent and the edge and bulk gaps are equal,
$\De_\text{edge}=\De_\text{bulk}=\De^\text{AF}$,
Fig.~\ref{fig:Ep}(a).
Upon increasing $\e_Z$, the edge gap $\De_\text{edge}=2\De_{zx}$ of the CAF phase gradually decreases as
$\sin \theta_s = \sqrt{1-(\frac{\e_Z}{2|u_\perp|})^2}$,
as the angle $2\theta_s$ between the spins decreases and
the CAF phase continuously transforms to the F phase,
Fig.~\ref{fig:Ep}(b).
The F phase is reached at $\e_Z = 2 |u_\perp|$
and persists upon further increase of $\e_Z > 2 |u_\perp|$;
the AF component turns zero and the edge gap closes, $\De_\text{edge}=\De_{zx}=0$, Fig.~\ref{fig:Ep}(c).
The F phase is characterized by the gapless counter-propagating edge excitations with opposite spin projections and spectra $\pm[\e(p)-\e_Z]$,
in accord with earlier findings~\cite{Abanin,FB,Gusynin}.

At the same time, according to Eq.~(\ref{eq:Debulk}),
the mean-field bulk gap $\De_\text{bulk}=\De^\text{AF}$ of the CAF phase, $\e_Z \leq 2 |u_\perp|$,
does not depend on the Zeeman energy $\e_Z$ and is equal to the gap (\ref{eq:DeAF}) of the AF phase.
In the F phase, $\e_Z> 2|u_\perp|$,  $\De_\text{bulk}$ grows with $\e_Z$ due to the Zeeman effect.

These findings explicitly confirm the expectation of Ref.~\onlinecite{MKBLG}
for the properties of the edge excitations of the CAF phase of the $\nu=0$ quantum Hall state~\cite{crudeness}.
The resulting physical behavior and experimental implications were discussed in the Introduction.

\section{Mean-field excitations of the canted antiferromagnetic phase in bilayer graphene\label{sec:BLG}}

The findings of Secs.~\ref{sec:model}-\ref{sec:excitations} for MLG are straightforwardly generalized to the case of BLG.
The key extra feature in BLG is that both $n=0$  and $n=1$ magnetic oscillator states belong the $\e=0$ LL~\cite{MF},
resulting in its additional two-fold degeneracy.
The bulk eigenstates $|p n\la\sig\ran$ of the $\e=0$ LL in the Landau gauge are characterized by momentum $p$,
valley $\la=K,K'$, spin $\sig=\ua,\da$, and $n=0,1$ quantum numbers. Each orbital $p$ is therefore eightfold-degenerate.

The projected Hamiltonian for the $\e=0$ LL in BLG, valid at energies $\e \ll 1/(m l_B^2)$ below the LL spacing
($m$ is the effective mass of the quadratic spectrum), has the form
\beq
    \Hh=\Hh_0+\Hh_{\itxt\circ} +\Hh_{\itxt\dm} + \Hh_Z,
\eeq
\beq
        \Hh_0 =
        %\pm \sum_p \e(p) c^\dg_{\la \sig, p} \tau^x_{\la\la'} c_{\la'\sig,p}=
        - \sum_{pn} \e_n(p) \ch^\dg_{pn} \Tc_x \ch_{pn},
\label{eq:H0BLG}
\eeq
\beq
    \Hh_{\itxt\circ} = \frac{1}{2} \sum_{p_1+p_1'=p_2+p_2'} g_0 \bar{V}^{n_1 p_1, n_2 p_2}_{n_1' p_1', n_2' p_2'}
     :\![\ch_{p_1 n_1}^\dg \ch_{p_2 n_2} ] [\ch_{p_1'n_1'}^\dg \ch_{p_2'n_2'}]\!:,
\label{eq:Hi0BLG}
\eeq
\beqar
    \Hh_{\itxt\dm} &=& \frac{1}{2} \sum_{\al=0,x,y,z} g_\al  \sum_{p_1+p_1'=p_2+p_2'} \bar{V}^{n_1 p_1, n_2 p_2}_{n_1' p_1', n_2' p_2'}\times
    \nonumber \\
        & & \times
     :\![\ch_{p_1 n_1}^\dg \Tc_\al \ch_{p_2 n_2} ] [\ch_{p_1'n_1'}^\dg \Tc_\al \ch_{p_2'n_2'}]\!:,
\label{eq:HidmBLG}
\eeqar
\beq
    \Hh_Z = - \e_Z\sum_{pn} \ch_{pn}^\dg S_z \ch_{pn},
\label{eq:HZBLG}
\eeq
\[
    \ch_{pn}=(\ch_{pn K\ua},\ch_{pn K\da},\ch_{pn K'\ua},\ch_{pn K'\da})^\ttxt.
\]

\begin{figure}
\centerline{\includegraphics[width=.35\textwidth]{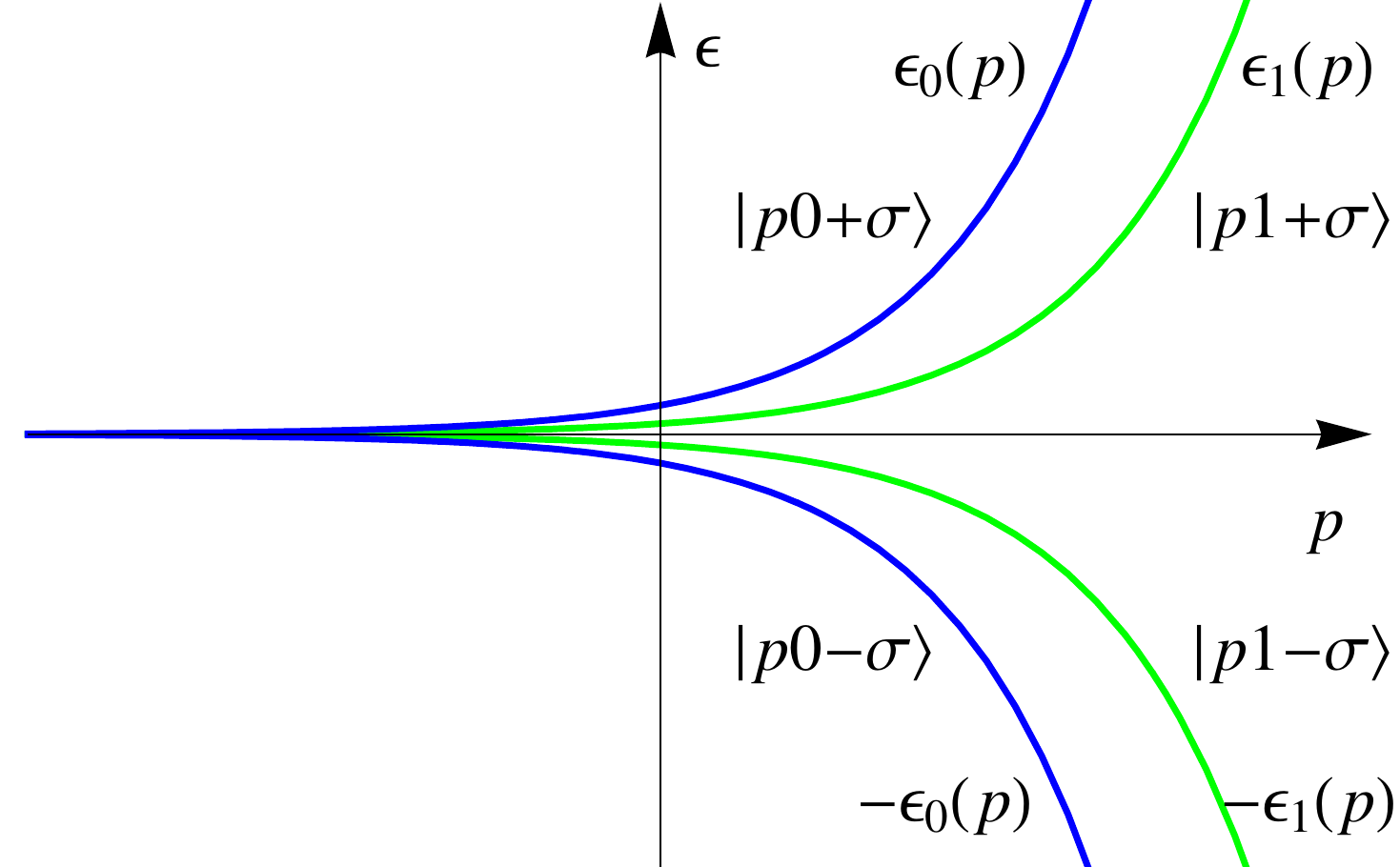}}
\caption{(Color online)
The structure of the single-particle spectrum pertaining to the $\e=0$ LL
in a BLG sample with ``armchair-like'' boundary, neglecting the Zeeman effect.
}
\label{fig:epBLG}
\end{figure}

\begin{figure*}
\centerline{\includegraphics[width=.33\textwidth]{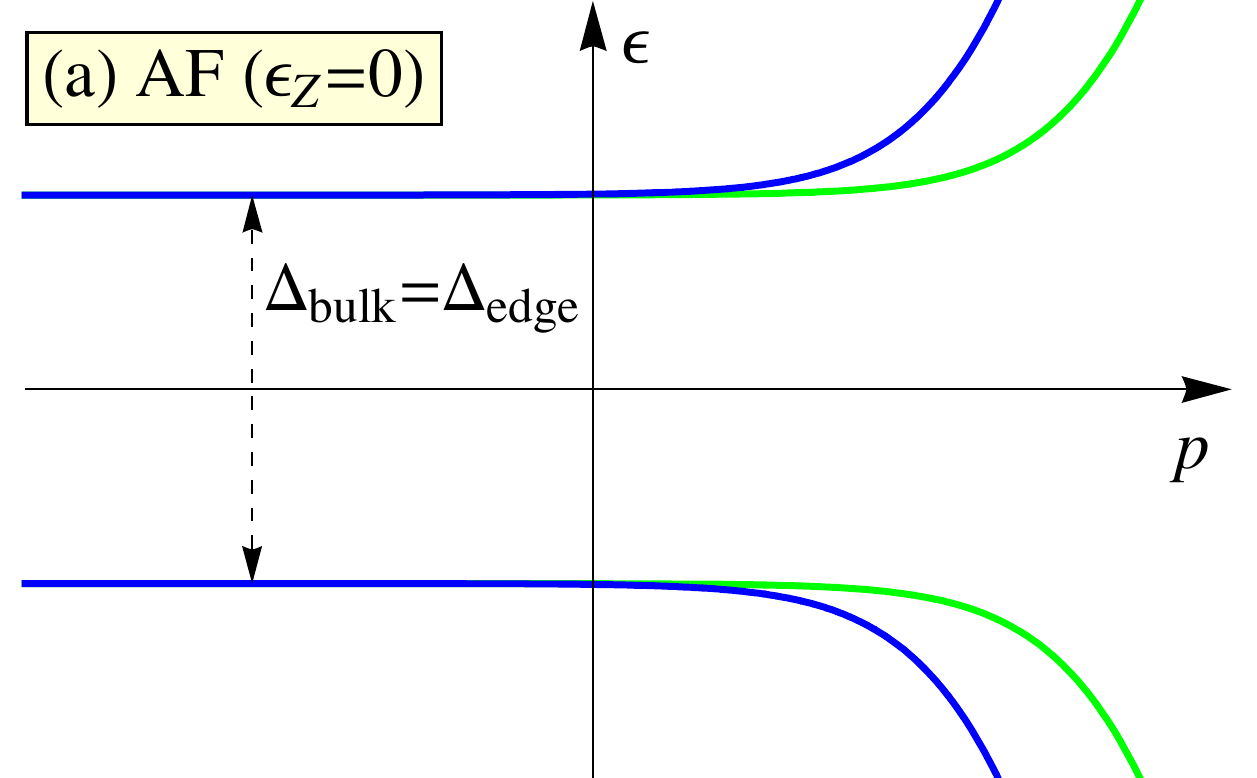}\includegraphics[width=.33\textwidth]{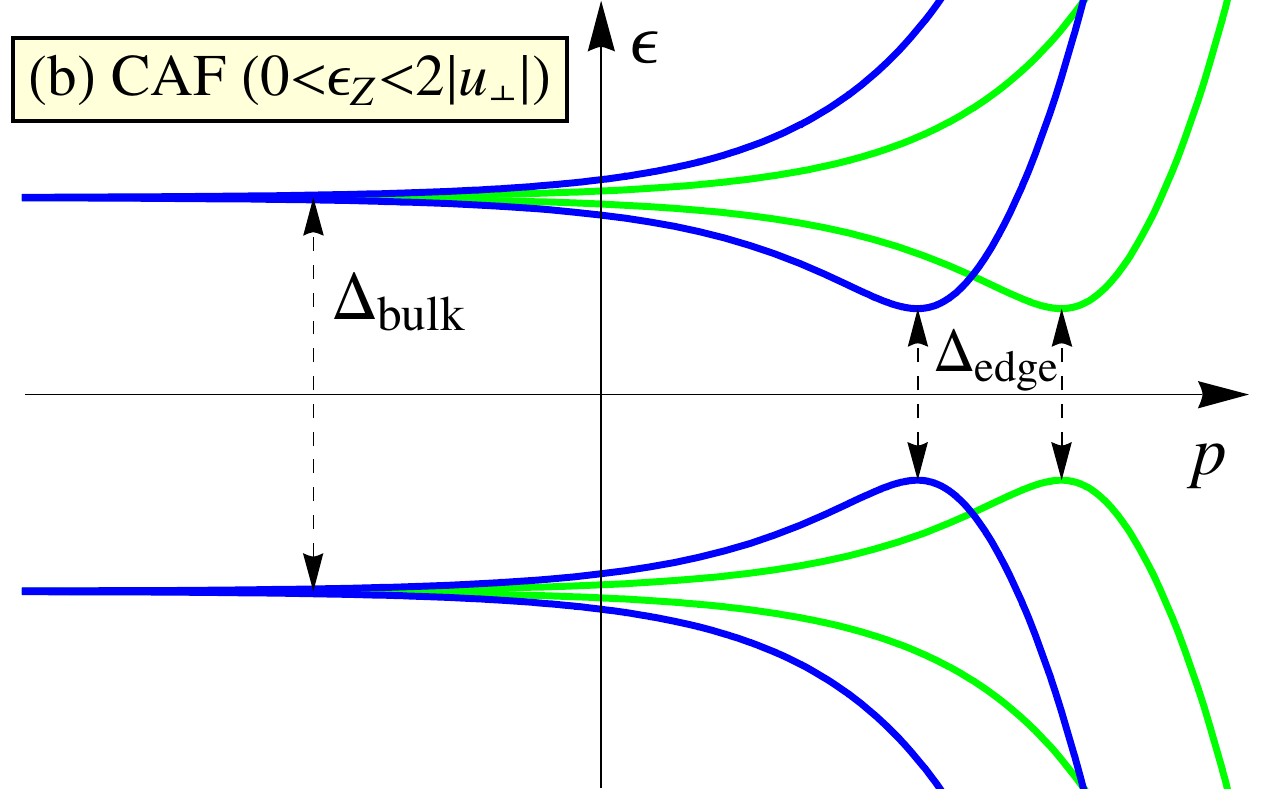}\includegraphics[width=.33\textwidth]{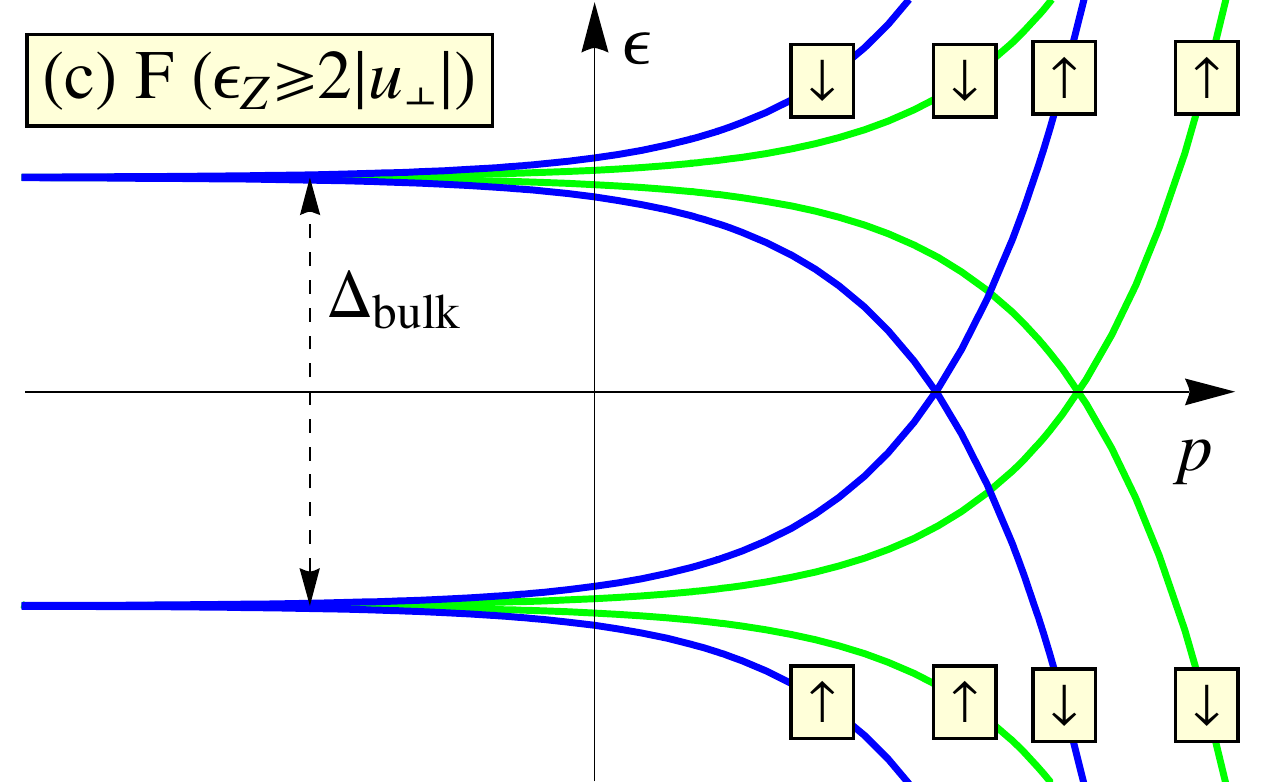}}
\caption{
(Color online)
The mean-field spectrum of the (a) antiferromagnetic (AF), (b) canted antiferromagnetic (CAF), and (c) ferromagnetic (F) phases of the $\nu=0$ state
of a bilayer graphene (BLG) sample with armchair-like boundary.
The spectrum (\ref{eq:EpBLG}) consists of eight branches $E_{n\pm\pm}(p)$  [Eq.~(\ref{eq:EpBLG})] --
two uncoupled sets of four branches pertaining to the $n=0$ (blue) and $n=1$ (green) states.
The edge $\De_\text{edge}$ [Eq.~(\ref{eq:Deedge})] and bulk $\De_\text{bulk}$ [Eq.~(\ref{eq:Debulk})]
gaps are the same for $n=0$ and $n=1$ sets and the same as in MLG;
their evolution with the Zeeman energy $\e_Z$ at a fixed anisotropy energy $u_\perp$
is presented in Fig.~\ref{fig:De}.
}
\label{fig:EpBLG}
\end{figure*}

Similarly to the considerations of Sec.~\ref{sec:edge}, at energies $\e \ll 1/(ml_B^2)$,
the effect of the edge of a BLG sample can be incorporated perturbatively
within the basis of the bulk eigenstates $|pn\la\sig\ran$ of the $\e=0$ LL.
For BLG, we also restrict ourselves to the case of ``armchair-like'' boundaries,
which, by definition, are described by an armchair boundary condition
in the two-band model of BLG~\cite{MF} and, consequently, do not contain dispersionless edge states in the absence of the magnetic field.
In this case, the LL spectrum has 8 branches pertaining to the $\e=0$ LL.
Deep in the bulk, where the kinetic energy vanishes,
the eight eigenstates $|pn\!\pm\!\sig \ran$
of the boundary problem {\em must} evolve into
the linear combinations of the bulk states $|pn\la\sig\ran$ that are related by the particle-hole symmetry,
\beq
    |pn\!\pm\!\sig\ran = \frac{1}{\sqrt{2}} (|pnK\sig\ran \mp |pnK'\sig\ran ) \mbox{  at  $\e_n(p)\rtarr 0$},
\label{eq:pmBLG}
\eeq
\beq
    c_{pn\pm\sig} = \frac{1}{\sqrt{2}} (c_{pnK\sig} \mp c_{pnK'\sig} ) \mbox{  at  $\e_n(p)\rtarr 0$}.
\label{eq:cpmBLG}
\eeq
This consideration simultaneously  proves that only the different valley states $|pnK\sig\ran$ and $|pnK'\sig\ran$ with the same $n$ and $\sig$
quantum numbers get hybridized by the edge.
Thus, the groups $|p0\la\sig\ran$  and $|p1\la\sig\ran$ of the bulk states evolve into two uncoupled sets of branches
with kinetic energies $ \pm \e_0(p)$ and $ \pm \e_1(p)$, as shown in Fig.~\ref{fig:epBLG},
which also justifies labeling the exact eigenstates $|pn\!\pm\!\sig\ran$ by the $n=0,1$ quantum numbers.
Since the $n=1$ states have a larger spatial extent than the $n=0$ states, the energies (arising from hybridization) 
are related as $\e_1(p)<\e_0(p)$.
The exact kinetic energy Hamiltonian for the states pertaining to the $\e=0$ LL reads
\beq
        \Hh_0 = \sum_{pn} \e_n(p) (c^\dg_{pn+\sig} c_{pn+\sig} - c^\dg_{pn-\sig} c_{pn-\sig}).
\label{eq:H0pmBLG}
\eeq
As in MLG, not too close to the edge, where $\e_n(p) \ll 1/(m l_B^2)$,
one may treat the kinetic energy (\ref{eq:H0pmBLG}) as a perturbation
and still use the relations (\ref{eq:pmBLG}) and (\ref{eq:cpmBLG}).
Substituting Eq.~(\ref{eq:pmBLG}) into Eq.~(\ref{eq:H0pmBLG}),
we arrive at the approximate expression (\ref{eq:H0BLG}) for kinetic energy Hamiltonian within the basis of the bulk eigenstates $|pn\la\sig\ran$.

The terms (\ref{eq:Hi0BLG}) and (\ref{eq:HidmBLG}) describe the valley-symmetric and asymmetric interactions, respectively.
Due to the difference in the orbital wave-functions of the $n=0$ and $n=1$ states, both interactions are
necessarily anisotropic in the 01-subspace. 
The valley-asymmetric interactions (\ref{eq:HidmBLG}) are point to a good approximation.
To keep the analysis simpler, 
we consider the valley-symmetric interactions (\ref{eq:Hi0BLG}) as point, as well,
in which case $u_0 =g_0/(\pi l_B^2)$ in the equations of Sec.~\ref{sec:excitations}.
Considering finite-range symmetric interactions would somewhat 
modify the spectrum quantitatively, but not qualitatively. 
The expression for the matrix element
$\bar{V}^{n_1 p_1, n_2 p_2}_{n_1' p_1', n_2' p_2'}$ 
is provided in the Appendix.

The technical steps leading to the edge excitations are analogous to the case of MLG.
One first obtains the bulk ground state of the $\nu=0$ state within the framework of QHFMism.
As demonstrated in Refs.~\onlinecite{Barlas,APS}, the asymmetry of the interactions in the 01-subspace
favors electrons to arrange into 01-singlet pairs with identical content
in the $KK'\otimes s$ space within a pair. Precisely, the states
\beq
    \Psi=  \prod_{pn}
    \lt(\sum_{\la\sig} \lan \la\sig|\chi_a\ran  c^\dg_{pn\la\sig} \rt)
    \lt(\sum_{\la'\sig'} \lan \la'\sig'|\chi_b\ran c^\dg_{pn\la'\sig'} \rt)|0\ran,
\label{eq:PsiB}
\eeq
deliver the variational minimum among the QHFM states in the full $01\otimes KK' \otimes s$ space
for the valley-symmetric interactions $\Hh_{\itxt\circ}$.
It is also possible to show that (\ref{eq:PsiB}) is, in fact, an eigenstate,
\[
    \Hh_{\itxt\circ}\Psi=E_0\Psi.
\]
However, the mutually orthogonal spinors $\chi_{a,b}$ in the $KK'\otimes s$ space
can still be arbitrary. Thus, at the level of the valley-symmetric interactions (\ref{eq:Hi0BLG}),
the (assumed) ground state $\Psi$ is degenerate.

To find the preferable order $P=\chi_a\chi_a^\dg +\chi_b \chi_b^\dg$
in the $KK'\otimes s$ space, one calculates the energy of the symmetry-breaking terms, the asymmetric interactions (\ref{eq:HidmBLG})
and Zeeman effect (\ref{eq:HZBLG}),
\begin{eqnarray}
    \Ec(P) &=& \Ec_\dm(P) + \Ec_Z(P),
\\
    \Ec_\dm(P) &=& \lan \Psi| \Hh_{\itxt \dm} |\Psi \ran/(2N),
\\
    \Ec_Z(P) &=& \lan \Psi| \Hh_Z |\Psi \ran/(2N).
\end{eqnarray}
With the definitions
\[
    u_\perp \equiv u_x=u_y =\frac{g_\perp}{\pi l_B^2}, \mbox{ }
    u_z =\frac{g_z}{\pi l_B^2},
\]
the functional dependencies of the isospin anisotropy  $\Ec_\dm(P)$ and Zeeman $\Ec_Z(P)$ energies on $P$
appear to be identical to those in MLG~\cite{MKBLG}, as given by Eqs.~(\ref{eq:Edm}) and (\ref{eq:EZ}).
Consequently, the phase diagram for the $\nu=0$ state in BLG, obtained by minimizing the energy $\Ec(P)$,
turns out formally identical~\cite{MKMLG,MKBLG} to that in MLG.

Owing to this correspondence,
the results for the excitations of the AF/CAF/F phases are basically the same, as well.
The mean-field Hamiltonian equals
\beq
    \Hh_\text{mf} \equiv \Hh_0+\Hh_\text{i,mf}+\Hh_Z = \sum_{pn} \ch^\dg_{pn} \hh_n \ch_{pn},
\label{eq:HmfBLG}
\eeq
\beq
    \hh_n = -\e_n(p) \tau_x \otimes \hat{1} - (\e_Z+\De_{0z}) \hat{1}\otimes \tau_z - \De_{zx} \tau_z \otimes \tau_x.
\label{eq:hBLG}
\eeq
Diagonalizing $\hh_n$ for each $n=0,1$,
we obtain the eight branches
\beq
    E_{n\pm\pm}(p) = \pm \sqrt{[\e_n(p) \pm (\e_Z+\De_{0z})]^2  + \De_{zx}^2},
\label{eq:EpBLG}
\eeq
of the mean-field spectrum of the AF/CAF/F phases of the $\nu=0$ state in BLG, shown in Fig.~\ref{fig:EpBLG}.
The condition (\ref{eq:thetas}) for the optimal angle $\theta_s$ of the AF/CAF/F phases,
the expressions for the bulk order parameter $P$ [Eq.~(\ref{eq:PCAF})], components $\De_{0z}$ [Eq.~(\ref{eq:De0z})] and $\De_{zx}$ [Eq.~\ref{eq:Dezx})]
of the mean-field potential $\Deh$ are the same as in MLG.

The only essential difference of BLG from MLG is that
due to the extra two-fold degeneracy of the $\e=0$ LL in BLG,
the number of branches is doubled:
there are two uncoupled sets, $n=0$ and $n=1$, of four branches
and for each set the spectrum (\ref{eq:EpBLG}) is the same as in MLG, compare with Eq.~(\ref{eq:Ep}).
The $n=0$ and $n=1$ sets
are not coupled by the interactions because
the order parameter $\hat{1}^{01} \otimes P$ of the state (\ref{eq:PsiB})
is singlet in the 01-subspace.

The edge $\De_\text{edge}$ [Eq.~(\ref{eq:Deedge})] and bulk $\De_\text{bulk}$ [Eq.~(\ref{eq:Debulk})]
gaps of the AF/CAF/F phases in BLG are the same as in MLG and the same for $n=0$ and $n=1$ sets, Fig.~\ref{fig:EpBLG}.
Also identical is the evolution of the $\De_\text{edge}$ and bulk $\De_\text{bulk}$ gaps
with varying $\e_Z/|u_\perp|$ by tilting the magnetic field, Fig.~\ref{fig:De}.
Since in the F phase ($\e_Z>2|u_\perp|$),
there are two gapless edge channels for each direction, $n=0$ and $n=1$, for negligible backscattering,
the two-terminal or Hall-bar longitudinal conductance
$G^\text{BLG} = 4e^2/h$ of the F phase
is quantized at twice the MLG values.
The resulting physical behavior and experimental implications
are thus essentially identical to those in MLG and were discussed in the Introduction.

\section{Conclusions\label{sec:conclusion}}

In conclusion, we studied the charge excitations of the canted antiferromagnetic phase
of the $\nu=0$ quantum Hall state in monolayer and bilayer graphene
within a simplified approach, namely,
neglecting the modification of the order parameter at the edge
and calculating the mean-field spectrum for a class of ``armchair-like'' boundaries.
We explicitly demonstrated that the gap in the edge spectrum of the canted antiferromagnetic phase
monotonically decreases upon decreasing the angle between the spin polarizations of the valleys=sublattices=layers,
as the canted antiferromagnetic phase continuously transforms to the fully spin-polarized phase,
Fig.~\ref{fig:CAF}, \ref{fig:Ep}, \ref{fig:De}, and \ref{fig:EpBLG}.
In practice such evolution is realized by tilting the magnetic field
and, in order to contrast this behavior to that of the competing spin-singlet phases,
the evolution with varying $B_\parallel$ and fixed $B_\perp$ should be traced.
The edge gap closes completely as the fully spin-polarized phase is reached
and gapless counter-propagating modes~\cite{Abanin,FB,Gusynin,JM} emerge.
This results in an gradual insulator-metal transition,
where the conductance grows exponentially in the insulating canted antiferromagnetic phase
and saturates to metallic values in the fully spin-polarized phase.

These unique edge transport properties of the canted antiferromagnetic phase,
earlier anticipated in Ref.~\onlinecite{MKBLG}, provide a straightforward way to identify
and distinguish it from other competing phases of the $\nu=0$ state in the experiment.
The data of the recent tilted-field experiment on bilayer graphene by Maher {\em et al.}~\cite{Maher} 
-- exponential growth of the longitudinal Hall-bar conductance $G_{xx}$
over a finite range of applied $B_\parallel$ at fixed $B_\perp$,
followed by the saturation to metallic values $G_{xx} \sim e^2/h$
-- are fully consistent with this predicted behavior, lending crucial support to the conclusion of Ref.~\onlinecite{MKBLG}
that the insulating $\nu=0$ quantum Hall state realized in real bilayer graphene~\cite{Weitz,Velasco}
is canted antiferromagnetic.

\section*{Acknowledgements}
Author is thankful to Patrick Maher, Andrea Young, and Philip Kim for insightful discussions of their unpublished experimental data~\cite{Maher}
and to Igor Aleiner for insightful discussions.
The research was supported by the U.S. DOE under contract DE-FG02-99ER45790
and in part by the National Science Foundation under grant No. NSF PHY11-25915.
The hospitality of the Kavli Institute for Theoretical Physics
program ``{\em The Physics of Graphene}'' where part of the work was performed is appreciated.

\section*{Appendix: Interaction matrix elements}

The interaction matrix elements in Eqs. (\ref{eq:Hi0}), (\ref{eq:Hidm}), (\ref{eq:Hi0BLG}), and (\ref{eq:HidmBLG})
are given by the standard expressions
\[
     V^{n_1, k+q_y/2, n_2, k-q_y/2}_{n_1',k'-q_y/2, n_2', k' + q_y/2 }=
\]
\beq
        = \frac{1}{L_y} \int \frac{\dt q_x}{2 \pi} \etxt^{\itxt q_x (k-k') l^2_B}
        K_{n_1 n_2}(\qb) K_{n_1' n_2'}(-\qb) V(q),
\label{eq:Vdef}
\eeq
for the conventional quadratic electron spectrum (since only the $\e=0$ LL states are involved).
Here, $n_1,n_2,n_1',n_2'$ are the LL indices, $V(q)$ is the interaction potential in the momentum space, $q=|\qb|$,
and
\beq
    K_{n n'}(\qb) = \int_{-\infty}^{+\infty} \dt x\,
    \etxt^{\itxt q_x x} \phi_n \lt(x- \frac{q_y}{2} l_B^2\rt ) \phi_{n'}\lt (x + \frac{q_y}{2} l_B^2\rt)
\label{eq:K}
\eeq
are the form-factors, with $\phi_n(x)$ the magnetic oscillator eigenstates.

In Eq.~(\ref{eq:Hi0}), $V^{p_1 p_2}_{p_1' p_2'}=V^{0p_1 0p_2}_{0p_1' 0p_2'}$ and
\[
    V(q) = \frac{V_0(q)}{1+\Pi(q)V_0(q)}
\]
is the screened Coulomb potential, with $V_0(q)=2\pi e^2/(\kappa q)$ and $\Pi(q)$ the static polarization operator.

In Eqs.~(\ref{eq:Hidm}), (\ref{eq:Hi0BLG}), and (\ref{eq:HidmBLG}), $\bar{V}^{p_1p_2}_{p_1'p_2'}=\bar{V}^{0p_1,0p_2}_{0p_1',0p_2'}$
and $V(q)=1$ is the point interaction potential of unit strength.


\begin{thebibliography}{99}
\bibitem{QHFMnote}
Here, ``ferromagnetism'' is understood in the generalized sense,
as ordering of whatever discrete degrees of freedom present in the system, and not necessarily just spin.
\bibitem{crudeness}
While we believe the qualitative behavior of the edge and bulk gaps [Eqs.~(\ref{eq:Deedge}) and (\ref{eq:Debulk}) and Fig.~\ref{fig:De}]
is correctly captured by the present approach,
we do not expect these dependencies to be necessarily quantitatively accurate,
due to the rather crude approximations involved.
A more reliable quantitative behavior can be anticipated from the generalization of the approach of Ref.~\onlinecite{FB}.
\end{thebibliography}
\end{document}